\documentclass[twocolumn]{aastex631}

\usepackage{verbatim}
\usepackage[usenames, dvipsnames]{}
\usepackage{xcolor}
\usepackage{threeparttable}
\usepackage{tabularx}
\usepackage{afterpage}

\newcommand{\C}{\ensuremath{^{\circ}C\;}}

\newcommand{\Mearth}{\ensuremath{M_\earth}}

\newcommand{\TESS}{TESS}

\newcommand{\bjdtdb}{\ensuremath{\rm {BJD_{TDB}}}}

\newcommand{\kms}{km\,s$^{-1}$}

\newcommand{\logg}{\ensuremath{\log g}}

\newcommand{\masyr}{\ensuremath{\rm mas\,yr^{-1}}}

\newcommand{\mj}{\ensuremath{\,M_{\rm J}}}

\newcommand{\ms}{\ensuremath{\rm m\,s^{-1}}}

\newcommand{\rj}{\ensuremath{\,R_{\rm J}}}

\newcommand{\rstar}{\ensuremath{R_\star}}

\newcommand{\Teff}{$T_{\rm eff}$}
\newcommand{\teff}{$T_{\rm eff}$}

\newcommand{\feh}{\ensuremath{\left[{\rm Fe}/{\rm H}\right]}}
\newcommand{\afe}{\ensuremath{\left[\alpha/{\rm Fe}\right]}}
\newcommand{\mh}{\ensuremath{\left[{\rm M}/{\rm H}\right]}}
\newcommand{\cfe}{\ensuremath{\left[{\rm C}/{\rm Fe}\right]}}
\newcommand{\nfe}{\ensuremath{\left[{\rm N}/{\rm Fe}\right]}}
\newcommand{\ofe}{\ensuremath{\left[{\rm O}/{\rm Fe}\right]}}
\newcommand{\nafe}{\ensuremath{\left[{\rm Na}/{\rm Fe}\right]}}
\newcommand{\mgfe}{\ensuremath{\left[{\rm Mg}/{\rm Fe}\right]}}
\newcommand{\alfe}{\ensuremath{\left[{\rm Al}/{\rm Fe}\right]}}
\newcommand{\sife}{\ensuremath{\left[{\rm Si}/{\rm Fe}\right]}}
\newcommand{\sfe}{\ensuremath{\left[{\rm S}/{\rm Fe}\right]}}
\newcommand{\cafe}{\ensuremath{\left[{\rm Ca}/{\rm Fe}\right]}}
\newcommand{\tife}{\ensuremath{\left[{\rm Ti}/{\rm Fe}\right]}}
\newcommand{\crfe}{\ensuremath{\left[{\rm Cr}/{\rm Fe}\right]}}
\newcommand{\mnfe}{\ensuremath{\left[{\rm Mn}/{\rm Fe}\right]}}
\newcommand{\cofe}{\ensuremath{\left[{\rm Co}/{\rm Fe}\right]}}
\newcommand{\nife}{\ensuremath{\left[{\rm Ni}/{\rm Fe}\right]}}
\newcommand{\cufe}{\ensuremath{\left[{\rm Cu}/{\rm Fe}\right]}}
\newcommand{\bafe}{\ensuremath{\left[{\rm Ba}/{\rm Fe}\right]}}

\newcommand{\thisstar}{TOI-4465}
\newcommand{\thisplanetb}{TOI-4465~b}
 
\newcommand{\plradunc}{1.25$^{+0.08}_{-0.07}$~\rj}

\newcommand{\plmassunc}{$5.89\pm0.26$~\mj}
\newcommand{\plrho}{3.73~g\,cm$^{-3}$} 
\newcommand{\plrhounc}{$3.73\pm0.53$~g\,cm$^{-3}$} 

\shorttitle{A Long-period super-Jupiter around TOI-4465}
\shortauthors{Essack et al.}
\submitjournal{\aj}

\begin{document}

\title{Giant Outer Transiting Exoplanet Mass (GOT `EM) Survey. VI: Confirmation of a Long-Period Giant Planet Discovered with a Single TESS Transit}

\author[0000-0002-2482-0180]{Zahra~Essack}
\affiliation{Department of Physics and Astronomy, The University of New Mexico, 210 Yale Blvd NE, Albuquerque, NM 87106, USA}


\author[0000-0003-2313-467X]{Diana~Dragomir}
\affiliation{Department of Physics and Astronomy, The University of New Mexico, 210 Yale Blvd NE, Albuquerque, NM 87106, USA}

\author[0000-0002-4297-5506]{Paul~A.~Dalba} 
\affiliation{Department of Astronomy and Astrophysics, University of California, Santa Cruz, CA 95064, USA}

\author[0000-0002-1357-9774]{Matthew~P.~Battley} 
\affiliation{Observatoire Astronomique de l’Université de Genève, Chemin Pegasi 51, CH-1290 Versoix, Switzerland}

\author[0000-0002-5741-3047]{David~ R.~Ciardi} 
\affiliation{Caltech/IPAC-NASA Exoplanet Science Institute, 770 S. Wilson Avenue, Pasadena, CA 91106, USA}

\author[0000-0001-6588-9574]{Karen~A.~Collins} 
\affiliation{Center for Astrophysics ${\rm \mid}$ Harvard {\rm \&} Smithsonian, 60 Garden Street, Cambridge, MA 02138, USA}

\author[0000-0002-2532-2853]{Steve~B.~Howell} 
\affiliation{NASA Ames Research Center, Moffett Field, CA 94035, USA}

\author{Matias~I.~Jones} 
\affiliation{European Southern Observatory, Alonso de Córdova 3107, Vitacura, Casilla 19001, Santiago, Chile}

\author[0000-0002-7084-0529]{Stephen~R.~Kane}
\affiliation{Department of Earth and Planetary Sciences, University of California, Riverside, CA 92521, USA}

\author[0000-0003-2008-1488]{Eric E. Mamajek} 
\affiliation{Jet Propulsion Laboratory, California Institute of Technology, 4800 Oak Grove Drive, Pasadena, CA 91109, USA}

\author[0000-0002-9312-0073]{Christopher R. Mann} 
\affiliation{National Research Council Canada, Herzberg Astronomy \& Astrophysics Research Centre, 5071 West Saanich Road, Victoria, BC V9E 2E7, Canada}

\author[0000-0002-4510-2268]{Ismael~Mireles} 
\affiliation{Department of Physics and Astronomy, The University of New Mexico, 210 Yale Blvd NE, Albuquerque, NM 87106, USA}

\author[0000-0002-2702-7700]{Dominic~Oddo} 
\affiliation{Department of Physics and Astronomy, The University of New Mexico, 210 Yale Blvd NE, Albuquerque, NM 87106, USA}

\author[0000-0001-6629-5399]{Lauren A. Sgro} 
\affiliation{SETI Institute, Carl Sagan Center, 339 Bernardo Ave, Suite 200, Mountain View, CA 94043, USA}

\author[0000-0002-3481-9052]{Keivan G.\ Stassun} 
\affiliation{Department of Physics and Astronomy, Vanderbilt University, Nashville, TN 37235, USA}

\author[0000-0003-2417-7006]{Solene~Ulmer-Moll} 
\affiliation{Observatoire Astronomique de l’Université de Genève, Chemin Pegasi 51, CH-1290 Versoix, Switzerland}
\affiliation{Space Research and Planetary Sciences, Physics Institute, University of Bern, Gesellschaftsstrasse 6, 3012 Bern, Switzerland}

\author[0000-0001-8621-6731]{Cristilyn N.\ Watkins}
\affiliation{Center for Astrophysics ${\rm \mid}$ Harvard {\rm \&} Smithsonian, 60 Garden Street, Cambridge, MA 02138, USA}

\author[0000-0001-7961-3907]{Samuel~W.~Yee} 
\affiliation{Center for Astrophysics ${\rm \mid}$ Harvard {\rm \&} Smithsonian, 60 Garden Street, Cambridge, MA 02138, USA}
\altaffiliation{51 Pegasi b Fellow}

\author[0000-0002-0619-7639]{Carl~Ziegler} 
\affiliation{Department of Physics, Engineering and Astronomy, Stephen F. Austin State University, 1936 North St, Nacogdoches, TX 75962, USA}

\author[0000-0001-6637-5401]{Allyson Bieryla} 
\affiliation{Center for Astrophysics ${\rm \mid}$ Harvard {\rm \&} Smithsonian, 60 Garden Street, Cambridge, MA 02138, USA}


\author[0009-0004-7473-4573]{Ioannis~Apergis} 
\affiliation{Department of Physics, University of Warwick, Gibbet Hill Road, Coventry CV4 7AL, UK}
\affiliation{Centre for Exoplanets and Habitability, University of Warwick, Gibbet Hill Road, Coventry CV4 7AL, UK}

\author[0000-0003-1464-9276]{Khalid Barkaoui}
\affiliation{Astrobiology Research Unit, Universit\'e de Li\`ege, 19C All\'ee du 6 Ao\^ut, 4000 Li\`ege, Belgium}
\affiliation{Department of Earth, Atmospheric and Planetary Science, Massachusetts Institute of Technology, 77 Massachusetts Avenue, Cambridge, MA 02139, USA}
\affiliation{Instituto de Astrof\'isica de Canarias (IAC), Calle V\'ia L\'actea s/n, 38200, La Laguna, Tenerife, Spain}

\author[0000-0002-9158-7315]{Rafael Brahm} 
\affiliation{Facultad de Ingeniera y Ciencias, Universidad Adolfo Ib\'{a}\~{n}ez, Av. Diagonal las Torres 2640, Pe\~{n}alol\'{e}n, Santiago, Chile}
\affiliation{Millennium Institute for Astrophysics, Chile}
\affiliation{Data Observatory Foundation, Chile}

\author[0000-0001-7904-4441]{Edward M. Bryant} 
\affiliation{Mullard Space Science Laboratory, University College London, Holmbury St Mary, Dorking, Surrey RH5 6NT, UK}

\author[0000-0002-0792-3719]{Thomas M. Esposito} 
\affiliation{SETI Institute, Carl Sagan Center, 339 Bernardo Ave, Suite 200, Mountain View, CA 94043, USA}
\affiliation{Unistellar, 5 all\'ee Marcel Leclerc, b\^{a}timent B, Marseille, 13008, France}
\affiliation{Department of Astronomy, University of California, Berkeley, CA 94720, USA}

\author{Pedro Figueira} 
\affiliation{Observatoire Astronomique de l’Université de Genève, Chemin Pegasi 51, CH-1290 Versoix, Switzerland}

\author[0000-0003-3504-5316]{Benjamin~J.~Fulton} 
\affiliation{NASA Exoplanet Science Institute/Caltech-IPAC, MC 314-6, 1200 E. California Blvd., Pasadena, CA 91125, USA}

\author[0000-0002-4259-0155]{Samuel~Gill} 
\affiliation{Department of Physics, University of Warwick, Gibbet Hill Road, Coventry CV4 7AL, UK}
\affiliation{Centre for Exoplanets and Habitability, University of Warwick, Gibbet Hill Road, Coventry CV4 7AL, UK}

\author[0000-0001-8638-0320]{Andrew~W.~Howard} 
\affiliation{Department of Astronomy, California Institute of Technology, Pasadena, CA 91125, USA}

\author[0000-0002-0531-1073]{Howard~Isaacson} 
\affiliation{Department of Astronomy, University of California Berkeley, Berkeley, CA 94720, USA}
\affiliation{Centre for Astrophysics, University of Southern Queensland, Toowoomba, QLD, Australia}

\author[0009-0006-0719-9229]{Alicia~Kendall} 
\affiliation{School of Physics and Astronomy, University of Leicester, University Road, Leicester, LE1 7RH, UK }

\author{Nicholas~Law} 
\affiliation{Department of Physics and Astronomy, The University of North Carolina at Chapel Hill, Chapel Hill, NC 27599-3255, USA}

\author[0000-0003-2527-1598]{Michael~B.~Lund} 
\affiliation{Caltech/IPAC-NASA Exoplanet Science Institute, 770 S. Wilson Avenue, Pasadena, CA 91106, USA}

\author[0000-0003-3654-1602]{Andrew~W.~Mann} 
\affiliation{Department of Physics and Astronomy, The University of North Carolina at Chapel Hill, Chapel Hill, NC 27599-3255, USA}

\author[0000-0001-7233-7508]{Rachel A. Matson} 
\affiliation{U.S. Naval Observatory, 3450 Massachusetts Avenue NW, Washington, D.C. 20392, USA}

\author[0000-0001-9087-1245]{Felipe Murgas}
\affiliation{Instituto de Astrof\'\i sica de Canarias (IAC), 38205 La Laguna, Tenerife, Spain}
\affiliation{Departamento de Astrof\'\i sica, Universidad de La Laguna (ULL), 38206, La Laguna, Tenerife, Spain}

\author[0000-0003-0987-1593]{Enric Palle}
\affiliation{Instituto de Astrof\'\i sica de Canarias (IAC), 38205 La Laguna, Tenerife, Spain}
\affiliation{Departamento de Astrof\'\i sica, Universidad de La Laguna (ULL), 38206, La Laguna, Tenerife, Spain}

\author[0000-0002-8964-8377]{Samuel N. Quinn} 
\affiliation{Center for Astrophysics ${\rm \mid}$ Harvard {\rm \&} Smithsonian, 60 Garden Street, Cambridge, MA 02138, USA}

\author[0000-0002-3844-1747]{Alexandre Revol} 
\affiliation{Observatoire Astronomique de l’Université de Genève, Chemin Pegasi 51, CH-1290 Versoix, Switzerland}

\author[0000-0001-8018-0264]{Suman~Saha} 
\affiliation{Instituto de Estudios Astrofísicos, Facultad de Ingeniería y Ciencias, Universidad Diego Portales, Av. Ejército Libertador 441, Santiago, Chile}
\affiliation{Centro de Excelencia en Astrofísica y Tecnologías Afines (CATA), Camino El Observatorio 1515, Las Condes, Santiago, Chile} 

\author[0000-0001-8227-1020]{Richard P. Schwarz}
\affiliation{Center for Astrophysics ${\rm \mid}$ Harvard {\rm \&} Smithsonian, 60 Garden Street, Cambridge, MA 02138, USA}

\author[0000-0003-3904-6754]{Ramotholo Sefako}  
\affiliation{South African Astronomical Observatory, P.O. Box 9, Observatory, Cape Town 7935, South Africa}

\author[0000-0002-1836-3120]{Avi Shporer}
\affiliation{Department of Physics and Kavli Institute for Astrophysics and Space Research, Massachusetts Institute of Technology, Cambridge, MA 02139, USA}

\author[0000-0003-0647-6133]{Ivan~A.~Strakhov} 
\affiliation{Sternberg Astronomical Institute Lomonosov Moscow State University Universitetskii prospekt, 13, Moscow 119992, Russia}

\author[0000-0001-6213-8804]{Steven~Villanueva,~Jr.}
\affiliation{NASA Goddard Space Flight Center, 8800 Greenbelt Road, Greenbelt, MD 20771, USA}


\author[0000-0003-2058-6662]{George~R.~Ricker}
\affiliation{Department of Physics and Kavli Institute for Astrophysics and Space Research, Massachusetts Institute of Technology, Cambridge, MA 02139, USA}

\author[0000-0001-6763-6562]{Roland~Vanderspek}
\affiliation{Department of Physics and Kavli Institute for Astrophysics and Space Research, Massachusetts Institute of Technology, Cambridge, MA 02139, USA}

\author[0000-0001-9911-7388]{David~W.~Latham}
\affiliation{Center for Astrophysics ${\rm \mid}$ Harvard {\rm \&} Smithsonian, 60 Garden Street, Cambridge, MA 02138, USA}

\author[0000-0002-6892-6948]{Sara~Seager}
\affiliation{Department of Physics and Kavli Institute for Astrophysics and Space Research, Massachusetts Institute of Technology, Cambridge, MA 02139, USA}
\affiliation{Department of Earth, Atmospheric and Planetary Sciences, Massachusetts Institute of Technology, Cambridge, MA 02139, USA}
\affiliation{Department of Aeronautics and Astronautics, MIT, 77 Massachusetts Avenue, Cambridge, MA 02139, USA}

\author[0000-0002-4265-047X]{Joshua~N.~Winn}
\affiliation{Department of Astrophysical Sciences, Princeton University, 4 Ivy Lane, Princeton, NJ 08544, USA}


\author[0000-0002-1514-5558]{Pau Bosch-Cabot}
\affiliation{Observatori Astronòmic Albanyà, Camí de Bassegoda S/N, Albanyà 17733, Girona, Spain}
\affiliation{University of Lethbridge, Lethbridge, AB, Canada}

\author[0000-0003-2781-3207]{Kevin I.\ Collins}
\affiliation{George Mason University, 4400 University Dr, Fairfax, VA 22030, USA}

\author[0000-0002-6482-2180]{Raquel For\'es-Toribio} \affiliation{Departamento de Astronom\'{\i}a y Astrof\'{\i}sica, Universidad de Valencia, E-46100 Burjassot, Valencia, Spain}
\affiliation{Observatorio Astron\'omico, Universidad de Valencia, E-46980 Paterna, Valencia, Spain}

\author{Fabian Rodriguez Frustagia}
\affiliation{American Association of Variable Star Observers, 49 Bay State Road, Cambridge, MA 02138, USA}
\affiliation{Frustaglia Private Observatory, Spain}

\author[0000-0002-5443-3640]{Eric Girardin}
\affiliation{Grand-Pra Observatory, 1984 Les Hauderes, Switzerland}

\author[0000-0002-0473-4437]{Ian J. Helm}
\affiliation{George Mason University, 4400 University Dr, Fairfax, VA 22030, USA}

\author[0000-0003-0828-6368]{Pablo Lewin}
\affiliation{The Maury Lewin Astronomical Observatory, Glendora,California.91741. USA}

\author[0000-0001-9833-2959]{Jose A. Mu\~noz} 
\affiliation{Departamento de Astronom\'{\i}a y Astrof\'{\i}sica, Universidad de Valencia, E-46100 Burjassot, Valencia, Spain}
\affiliation{Observatorio Astron\'omico, Universidad de Valencia, E-46980 Paterna, Valencia, Spain}

\author{Patrick Newman}
\affiliation{George Mason University, 4400 University Dr, Fairfax, VA 22030, USA}

\author[0000-0002-8864-1667]{Peter Plavchan}
\affiliation{George Mason University, 4400 University Dr, Fairfax, VA 22030, USA}

\author{Gregor Srdoc}
\affil{Kotizarovci Observatory, Sarsoni 90, 51216 Viskovo, Croatia} 

\author[0000-0003-2163-1437]{Chris Stockdale}
\affiliation{Hazelwood Observatory, Australia}

\author[0000-0002-6176-9847]{Ana\"el W\"unsche}
\affiliation{Laboratoire d'Astrophysique des Baronnies, Science Citoyenne et Actions pour la Nuit (LABSCAN), Moydans, France}


\author[0000-0002-3278-9590]{Mario Billiani}
\affiliation{Unistellar Citizen Scientist, Vienna, Austria}

\author{Martin Davy}
\affiliation{Unistellar Citizen Scientist, Maxéville, France}

\author[0009-0009-8377-0695]{Alex Douvas}
\affiliation{Unistellar Citizen Scientist, Temecula, California, US}

\author[0000-0002-9297-5133]{Keiichi Fukui}
\affiliation{Unistellar Citizen Scientist, Tsuchiura, Ibaraki, Japan}

\author[0000-0003-4091-0247]{Bruno Guillet} 
\affiliation{Unistellar Citizen Scientist, Puy-Saint-Vincent, France}

\author{Cory Ostrem}
\affiliation{Unistellar Citizen Scientist, Ames, Iowa, US}

\author{Michael Rushton}
\affiliation{Unistellar Citizen Scientist, Tonbridge, Kent, England}

\author{Angsar Schmidt}
\affiliation{Unistellar Citizen Scientist, Archenhold Observatory, Finowfurt, Brandenburg, Germany}


\author{Andrea Finardi} 
\affiliation{Unistellar Citizen Scientist, Valeggio sul Mincio, Verona, Italy}

\author[0009-0004-8835-6059]{Patrice Girard} 
\affiliation{Unistellar Citizen Scientist, Bayonne, France}

\author[0000-0002-9540-6112]{Tateki Goto} 
\affiliation{Unistellar Citizen Scientist, Tonosyo, Kagawa, Japan}

\author[0000-0003-3464-1554]{Julien S. de Lambilly} 
\affiliation{Unistellar Citizen Scientist, Renens, Vaud, Switzerland}

\author[0000-0003-3046-9187]{Liouba Leroux}
\affiliation{Unistellar Citizen Scientist, Lattes, France}

\author[0000-0002-6818-6599]{Fabrice Mortecrette}
\affiliation{Unistellar Citizen Scientist, Normandie, France}

\author[0000-0001-9475-0344]{John W. Pickering} 
\affiliation{Unistellar Citizen Scientist, Christchurch, New Zealand}

\author[0000-0003-3462-7533]{Michael Primm} 
\affiliation{Unistellar Citizen Scientist, Austin, Texas, US}

\author{Marc Ribot} 
\affiliation{Unistellar Citizen Scientist, Saint Paul les Dax, France}

\author{Ethan Teng} 
\affiliation{Unistellar Citizen Scientist, Oakland, California, US}

\author{Aad Verveen} 
\affiliation{Unistellar Citizen Scientist, Laren, Gelderland, Netherlands}

\author[0000-0003-0404-6279]{Stefan Will}
\affiliation{Unistellar Citizen Scientist, Raleigh, North Carolina, US}

\author{Mark Ziegler} 
\affiliation{Unistellar Citizen Scientist, Longmont, Colorado, US}

\correspondingauthor{Zahra Essack}
\email{zessack@unm.edu}

\begin{abstract}
We report the discovery and confirmation of \thisplanetb, a \plradunc, \plmassunc\ giant planet orbiting a G dwarf star at $d\simeq$ 122~pc. The planet was detected as a single-transit event in data from Sector~40 of the Transiting Exoplanet Survey Satellite (TESS) mission. Radial velocity (RV) observations of TOI-4465 showed a planetary signal with an orbital period of $\sim$102~days, and an orbital eccentricity of $e=0.24\pm0.01$. TESS re-observed \thisstar\ in Sector 53 and Sector 80, but did not detect another transit of \thisplanetb\, as the planet was not expected to transit during these observations based on the RV period. A global ground-based photometry campaign was initiated to observe another transit of \thisplanetb\ after the RV period determination. The $\sim$12 hour-long transit event was captured from multiple sites around the world, and included observations from 24 citizen scientists, confirming the orbital period as $\sim$102~days. \thisplanetb\ is a relatively dense (\plrhounc), temperate (375--478~K) giant planet. Based on giant planet structure models, \thisplanetb\ appears to be enriched in heavy elements at a level consistent with late-stage accretion of icy planetesimals. Additionally, we explore TOI-4465~b's potential for atmospheric characterization, and obliquity measurement. Increasing the number of long-period planets by confirming single-transit events is crucial for understanding the frequency and demographics of planet populations in the outer regions of planetary systems.
\end{abstract}



\section{Introduction \label{sec:intro}}

Long-period exoplanets are important members of the broader exoplanet population. Warm giant planets (or warm Jupiters) are typically defined as gas giant planets with orbital periods of 10-200 days. Unlike their close-in hot Jupiter counterparts, the moderate temperatures of warm Jupiters allow for the investigation of gas giant atmospheres under less extreme conditions, where intense stellar irradiation does not significantly alter the planetary radius, atmospheric chemistry, or dynamics \citep[e.g.][]{guillot2002evolution, fortney2010interior}. Longer-period planets experience significantly less atmospheric mass loss driven by stellar insolation than their shorter-period counterparts, and are more likely to retain their primordial compositions \citep[e.g.][]{fortney2007planetary, kubyshkina2019kepler}. Warm Jupiters and longer-period gas giants can serve as a link between the extreme hot Jupiters and cold solar system gas giants, allowing us to probe the relatively underexplored atmospheric physics and chemistry at cooler temperatures.

Warm giant planets can also provide insights into the formation and migration processes of planetary systems. Unlike hot Jupiters, which are strongly affected by stellar irradiation and tidal interactions \citep[e.g.][]{albrecht2012obliquities}, warm Jupiters allow for a more direct comparison of their properties to those predicted by different formation and migration scenarios \citep[e.g.][]{huang2016warm, rice2022tendency}. Hot Jupiter origin theories generally fall into two categories: high-eccentricity migration \citep[e.g.][]{rasio1996dynamical, fabrycky2007shrinking, wu2011secular}, and disk migration \citep[e.g.][]{goldreich1980disk, ward1997protoplanet}. Studying these intermediate warm Jupiter systems can help differentiate between these models by analyzing orbital eccentricities, obliquities, and planetary compositions.

The Transiting Exoplanet Survey Satellite (TESS; \citealp{ricker2015transiting}), launched in 2018, has been instrumental in expanding the exoplanet catalog through an all-sky survey of bright, nearby stars. With its wide field of view and high photometric precision, TESS has successfully detected numerous short-period exoplanets. However, TESS's typical observational baseline of 27~days limits its ability to detect long-period planets, which may only transit once during a TESS sector, resulting in single-transit events \citep[e.g.][]{villanueva2019estimate, cooke2021resolving}. 

Confirming and characterizing long-period planets from such single-transit events requires extensive radial velocity (RV) and/or photometric monitoring campaigns. The large orbital distances of long-period planets reduces their transit probabilities \citep{winn2010exoplanet}, and poses challenges for ground-based follow-up observations. As a result of these observational challenges, longer-period planets are underrepresented within the exoplanet catalog. Efforts to build a robust sample of long-period planets are crucial for understanding the frequency and demographics of planet populations in the outer regions of planetary systems.

In this paper, we report the discovery and confirmation of a long-period, massive, temperate giant planet transiting the G dwarf star TOI-4465, and initially detected as a TESS single-transit event. In Section \ref{sec:stellardata}, we present observations and characterization of the host star. In Section \ref{sec:planetdata}, we describe the time-series photometry and RV datasets obtained for the TOI-4465 system. In Section \ref{sec:juliet_jointfit}, we describe our data analysis, and derive properties for the planetary system. In Section \ref{sec:discussion}, we discuss the newly confirmed planet, TOI-4465~b, in the context of the wider exoplanet population, and present avenues for further characterization of the system. Finally, we present our conclusions in Section \ref{sec:conclusions}.


\section{Stellar Data \& Characterization \label{sec:stellardata}}

The astrometric, kinematic, photometric, and bulk properties of TOI-4465 (TIC 157236902) are summarized in Table \ref{tab:stellar_properties}. TOI-4465 appears in numerous stellar catalogs, but the previously anonymous star in the Hercules constellation has not been discussed in detail in any publications.

Previously published effective temperature (\teff) and surface gravity (\logg) estimates are summarized in Table \ref{tab:stellar_teff}.
Besides Gaia, the only previous spectral characterization came from data releases in the LAMOST survey \citep[e.g.][]{Luo2016}.
The previously published values strongly cluster near
\teff\, $\simeq$ 5540\,K ($\pm150$\,K rms scatter) and \logg\, $\simeq$ $4.35$ ($\pm0.1$ rms scatter), broadly
consistent with typical parameters for a G7V star, albeit with slightly lower surface gravity \citep[\logg\, $\simeq$ 4.45 is typical for main sequence stars of this \Teff\, and within $\pm$0.1 dex of solar metallicity;][]{Soubiran2022}. 

\begin{center}
\begin{deluxetable*}{lccc}
\tabletypesize{\footnotesize}
\tablecolumns{4}
\tablewidth{0pt}
\tablecaption{Astrometry, Photometry \& Properties for TOI-4465 \label{tab:stellar_properties}}
\tablehead{
\colhead{Parameter} & 
\colhead{Value} & 
\colhead{Unit} & 
\colhead{Ref.}
}
\startdata
\multicolumn{2}{l}{Identifying Information:}& \\
TESS ID & TIC 157236902 & $-$ & TESS\\
TYC ID & TYC 1569-1212-1 & $-$ & \emph{Tycho}\\
2MASS ID & 2MASS J18243222+1550301 & $-$ & 2MASS\\
GAIA ID & 4521842556801236352 & $-$ & \emph{Gaia} DR3\\
\hline
\multicolumn{2}{l}{Astrometric Properties:}& \\
$\alpha_{ICRS}$ & 276.13402341451 & deg & \citet{GaiaDR3}\\
$\delta_{ICRS}$ & +15.84189162349 & deg & \citet{GaiaDR3}\\
$\varpi$ & $8.1756\pm0.0197$ & mas & \citet{GaiaDR3}\\
$d$ & $121.931^{+0.287}_{-0.285}$ & pc & \citet{BailerJones2021}\\
$\mu_{\alpha}$ & $-54.282\pm0.017$ & \masyr & \citet{GaiaDR3}\\
$\mu_{\delta}$ & $43.706\pm0.018$ & \masyr & \citet{GaiaDR3}\\
$v_r$ & $-49.66\pm0.20$ & \kms & \citet{GaiaDR3}\\
$U$ & $-47.59\pm0.15$ & \kms & \citet{GCNS2021}\\
$V$ & $-32.80\pm0.14$ & \kms & \citet{GCNS2021}\\
$W$ & $27.10\pm0.10$ & \kms & \citet{GCNS2021}\\
\hline
\multicolumn{2}{l}{Magnitudes:}& \\
$U$ & 11.465 & mag & \citet{GaiaDR3}\\
$B$ & 11.229 & mag & \citet{GaiaDR3}\\
$V$ & 10.514 & mag & \citet{GaiaDR3}\\
$V$ & $10.497\pm0.018$ & mag & \citet{Henden2012}\\
$R_c$ & 10.105 & mag & \citet{GaiaDR3}\\
$I_c$ & 9.724 & mag & \citet{GaiaDR3}\\
$g$ & $10.825\pm0.061$ & mag & \citet{Henden2012}\\
$r$ & $10.315\pm0.041$ & mag & \citet{Henden2012}\\
$i$ & $10.146\pm0.049$ & mag & \citet{Henden2012}\\
$z$ & $10.062\pm0.041$ & mag & \citet{Henden2012}\\
$G$ & $10.328\pm0.003$ & mag & \citet{GaiaDR3}\\
$Bp-Rp$ & 0.897924 & mag & \citet{GaiaDR3}\\
$Bp-G$ & 0.364063 & mag & \citet{GaiaDR3}\\
$G-Rp$ & 0.533861 & mag & \citet{GaiaDR3}\\
$J$ & $9.191\pm0.020$ & mag & \citet{Cutri2003}\\
$H$ & $8.876\pm0.021$ & mag & \citet{Cutri2003}\\
$Ks$ & $8.787\pm0.016$ & mag & \citet{Cutri2003}\\
$W1$ & $8.736\pm0.023$ & mag & \citet{Cutri2012}\\
$W2$ & $8.785\pm0.021$ & mag & \citet{Cutri2012}\\
$W3$ & $8.752\pm0.025$ & mag & \citet{Cutri2012}\\
$W4$ & $8.865\pm0.379$ & mag & \citet{Cutri2012}\\
\hline
\multicolumn{2}{l}{Bulk Properties:}& \\
$R_\star$ & $1.014 \pm 0.041$ & R$_\odot$ & Section \ref{sec:SED}\\
$M_\star$ & $0.93 \pm 0.06$ & M$_\odot$ & Section \ref{sec:SED}\\
$T_\mathrm{eff}$ & $5545 \pm 19$ & K & \citet{Luo2022} (LAMOST DR7)\\
$\mathrm{[Fe/H]}$ & $-0.06 \pm 0.016$ & dex & \citet{Luo2022} (LAMOST DR7)\\
\logg & $4.379\pm0.030$ & dex & \citet{Luo2022} (LAMOST DR7)\\
$v$sin$i$ & $< 4.13$ & \kms & Section \ref{subsubsec:coralie} \\
\enddata
\tablecomments{
$\alpha_{ICRS}$ and $\delta_{ICRS}$ positions are for epoch 2016.0 from \citet{GaiaDR3}. Values from the APASS SUrvey \citet{Henden2012} are from Data Release 10 (\url{https://www.aavso.org/download-apass-data}).}
\end{deluxetable*}
\end{center}

\subsection{Spectral Energy Distribution \label{sec:SED}}

\begin{figure}
\centering
\includegraphics[width=\linewidth,trim=80 70 50 50,clip]{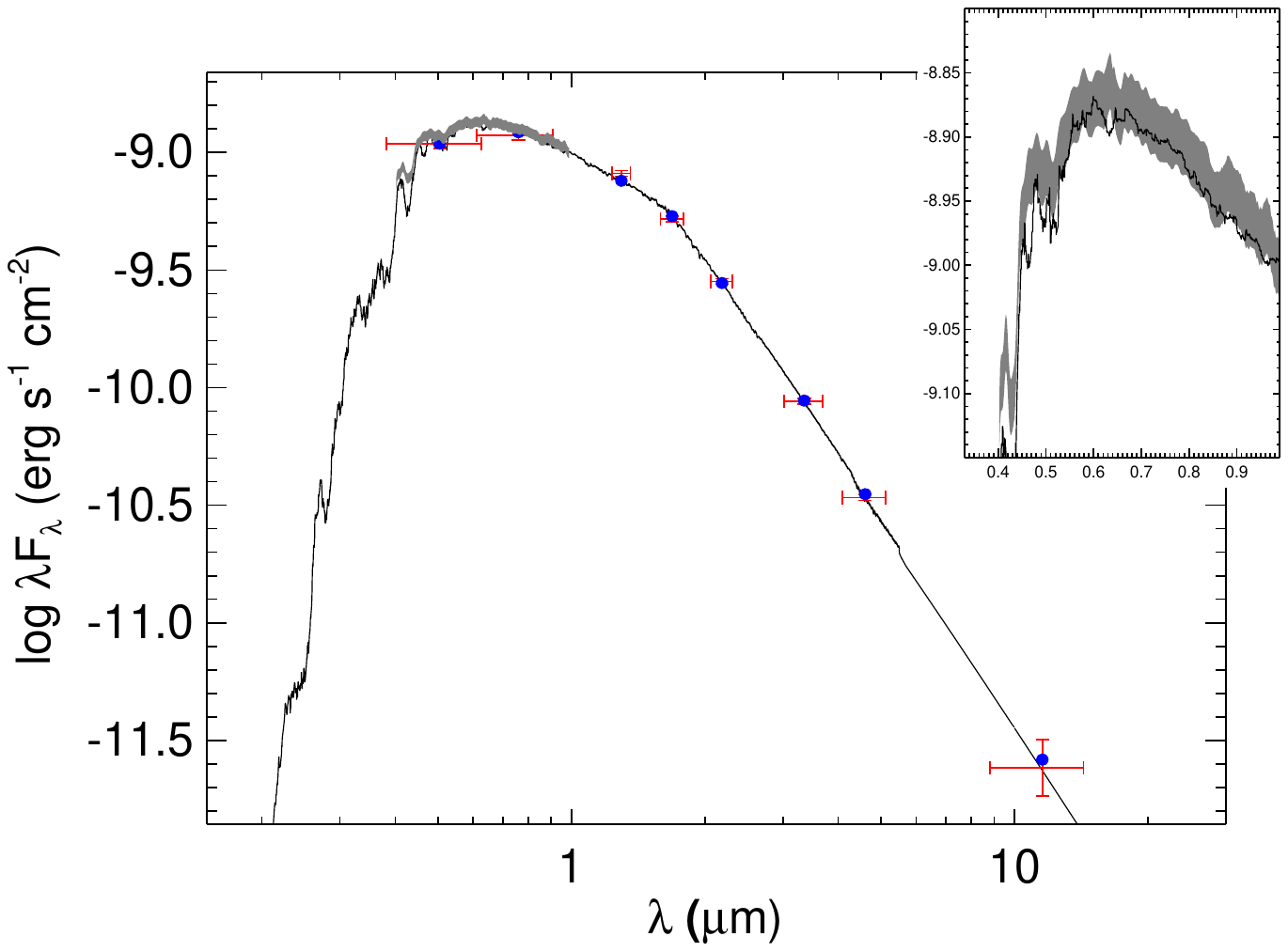}
\caption{Spectral energy distribution of TOI-4465. Red symbols represent the observed photometric measurements, where the horizontal bars represent the effective width of the passband. Blue symbols are the model fluxes from the best-fit PHOENIX atmosphere model (black). The inset shows the absolute flux-calibrated {\it Gaia\/} spectrophotometry as a gray swathe overlaid on the model (black). \label{fig:sed}}
\end{figure}

As an independent determination of the basic stellar parameters, we performed an analysis of the broadband spectral energy distribution (SED) of the star together with the {\it Gaia\/} DR3 parallax \citep[with no systematic offset applied; see, e.g.,][]{StassunTorres:2021}, in order to determine an empirical measurement of the stellar radius, following the procedures described in \citet{Stassun:2016}, \citet{Stassun:2017}, and \citet{Stassun:2018}. We obtained the $JHK_S$ magnitudes from {\it 2MASS}, the W1--W3 magnitudes from {\it WISE}, the $G_{\rm BP}$ and $G_{\rm RP}$ magnitudes from {\it Gaia}, and the absolute flux-calibrated {\it Gaia\/} spectrophotometry. The previously published Johnson $V$ magnitude with the smallest quoted uncertainties is $V = 10.497\pm0.018$ mag from 
AAVSO Photometric All-Sky Survey (APASS) Data Release 10\footnote{\href{https://www.aavso.org/download-apass-data}{https://www.aavso.org/download-apass-data}. Survey described in \citet{Henden2012}.}, consistent with the synthetic estimate from Gaia DR3
\citep[$V = 10.514$ mag;][]{GaiaDR3}. 
Together, the available photometry spans the full stellar SED over the wavelength range 0.4--10~$\mu$m (see Figure~\ref{fig:sed}). 

We performed a fit using PHOENIX stellar atmosphere models \citep{Husser:2013}, with the effective temperature ($T_{\rm eff}$), surface gravity ($\log g$), and metallicity ([Fe/H]) set to the spectroscopically determined values. The extinction $A_V$ was limited to the maximum line-of-sight value from the Galactic dust maps of \citet{Schlegel:1998}. The resulting fit (Figure~\ref{fig:sed}) has a best-fit $A_V = 0.01 \pm 0.01$ with a reduced $\chi^2$ of 1.5.
Based on the reddening curves of \citet{McCall2004} and \citet{Zhang2023}, this extinction is consistent with reddening E($B-V$) = $0.003\pm0.003$ mag, Gaia $G$-band extinction of $A_G$ = $0.009\pm0.009$ mag, and
2MASS $K_s$-band extinction $A_{Ks}$ = $0.0009\pm0.0009$ mag.
Combining these extinction estimates with the Gaia DR3 parallax and
the $V$, $G$, and $K_s$ magnitudes from APASS DR10, Gaia DR3, and 2MASS, respectively, results in absolute magnitude estimates of
$M_V$ = $5.05\pm0.02$, $M_G$ = $4.88\pm0.01$, and $M_{Ks}$ = 
$3.35\pm0.02$. Integrating the (unreddened) model SED gives the bolometric flux at Earth, $F_{\rm bol} = 1.765\pm0.020 \times 10^{-9}$ erg~s$^{-1}$~cm$^{-2}$ (apparent bolometric magnitude $m_{bol}$ = $10.386\pm0.012$ mag on the IAU 2015 scale\footnote{\url{https://www.iau.org/static/resolutions/IAU2015_English.pdf}}). 
Taking the $F_{\rm bol}$ and {\it Gaia\/} parallax directly gives the bolometric luminosity, $L_{\rm bol}$ = $0.8234\pm0.0097$~$L_{\odot}$ ($M_{bol}$ = $4.948\pm0.013$ mag). The Stefan-Boltzmann relation then yields the stellar radius, $R_\star = 1.014 \pm 0.041$~R$_\odot$. In addition, we can estimate the stellar mass from the empirical relations of \citet{Torres:2010}, giving $M_\star = 0.93 \pm 0.06$~M$_\odot$. 

Finally, we may estimate the projected rotation period from the spectroscopic $v\sin i$ together with \rstar, giving $P_{\rm rot}/\sin i = 12.8 \pm 6.4$~d, which from the empirical gyrochronology relations of \citet{Mamajek:2008} gives an estimated stellar age of $0.9 \pm 0.7$~Gyr. Since the spectroscopic $v\sin i$ is an upper limit estimated from the CORALIE RV observations (Section \ref{subsubsec:coralie}), the stellar age estimated from gyrochronology relations is a lower limit. 

For the subsequent analyses in this paper, we adopt the stellar radius and mass derived from the SED analysis, namely $R_\star = 1.014 \pm 0.041$~R$_\odot$, and $M_\star = 0.93 \pm 0.06$~M$_\odot$.

\begin{center}
\begin{deluxetable*}{ccc}
\tablecolumns{4}
\tablewidth{0pt}
\tablecaption{Previously Published Effective Temperature and Surface Gravity Estimates for \thisstar\label{tab:stellar_teff}}
\tablehead{
\colhead{\Teff} & 
\colhead{\logg} & 
\colhead{Ref.}}
\startdata
$5267^{+27}_{-46}$ & $4.20^{+0.11}_{-0.10}$ & GaiaDR3 GSP-Spec ANN using RVS spectra\\
$5440^{+380}_{-120}$ & ... & \citet{GaiaDR2}\\
$5443^{+4}_{-3}$ & $4.388\pm0.0002$ & GaiaDR3 GSP-Phot Aeneas best library using BP/RP spectra\\
$5503\pm20$ & $4.299\pm0.039$ & \citet{Xiang2019} (LAMOST DR5)\\
$5523\pm20$ & $4.344\pm0.031$ & \citet{Luo2019} (LAMOST DR5)\\	
$5534\pm1$ & $4.386\pm0.001$ & \citet{Queiroz2020}\\
$5545\pm19$ & $4.379\pm0.030$ & \citet{Luo2022} (LAMOST DR7)\\
$5557^{+176}_{-82}$ & $4.436^{+0.065}_{-0.094}$ & \citet{Stassun2019} (TIC V8.0)\\
$5616^{+23}_{-27}$ & $4.17^{+0.08}_{-0.06}$ & GaiaDR3 GSP-Spec MatisseGauguin using RVS spectra\\
$5629$ & 4.257 & \citet{Kordopatis2023}\\
$5783^{+151}_{-167}$ & $4.42^{+0.05}_{-0.01}$ & \citet{Anders2022}\\
$5851^{+294}_{-239}$ & $4.445^{+0.042}_{-0.040}$ & \citet{Anders2019}\\ 
\enddata
\tablecomments{
Effective temperatures and uncertainties rounded to nearest Kelvin, and surface gravities quoted to nearest 0.001 dex when uncertainties are reported to overprecision.
}
\end{deluxetable*}
\end{center}

\subsection{High Resolution Imaging} 

As part of our standard process for validating transiting exoplanets to assess the possible contamination of bound or unbound companions on the derived planetary radii \citep{ciardi2015}, we observed TOI~4465 with optical speckle observations at Gemini, SOAR, and SAI, and near-infrared adaptive optics (AO) imaging at Palomar and Lick Observatories.

\subsubsection{Optical Speckle Imaging}

Close stellar companions (bound or line of sight) can obscure a detected exoplanet transit signal, potentially leading to false positives. Even real planet discoveries may result in incorrect stellar and exoplanet parameters if a close companion exists and is unaccounted for \citep{ciardi2015, furlan2017densities, furlan2020unresolved}. Additionally, the presence of a close companion star can prevent the detection of small planets within the same exoplanetary system \citep{lester2021speckle}. Given that nearly one-half of solar-type stars are in binary or multiple star systems \citep{matson2018stellar}, we use high-resolution imaging to search for close companions unresolved in other follow-up observations. High-resolution imaging provides crucial information toward our understanding of exoplanetary formation, dynamics and evolution \citep{howell2021nasa}. 

\begin{figure}
    \centering
    \includegraphics[width=\linewidth]{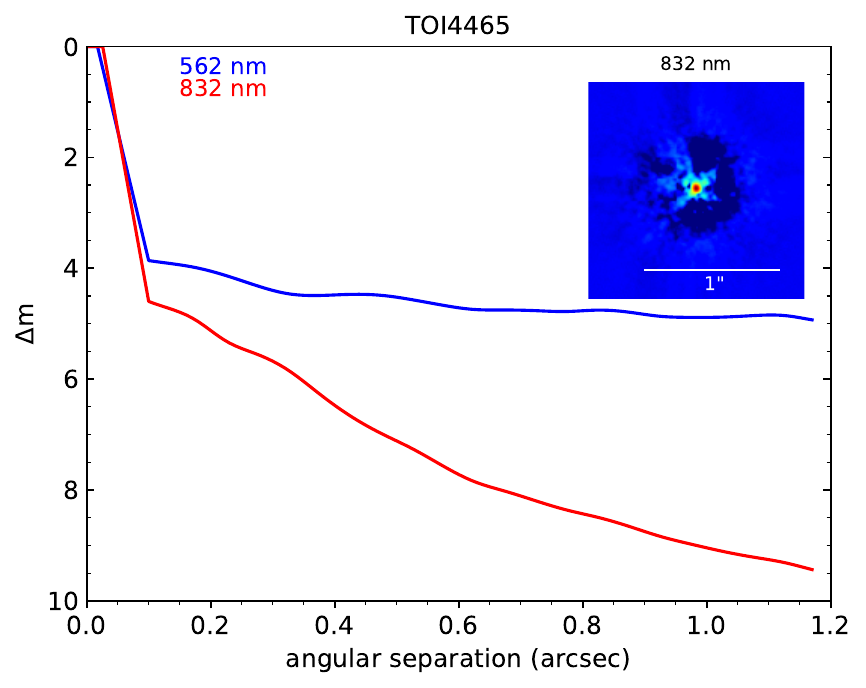}
\caption{Gemini-North ‘Alopeke $5\sigma$ speckle imaging contrast curves in 562~nm filter (blue line) and 832~nm filter (red line) as a function of the angular separation out to 1.2”. The inset shows the reconstructed 832 nm image of TOI-4465 with a 1” scale bar. TOI-4465 was found to have no close companions from the diffraction limit out to 1.2” to within the contrast levels achieved.  \label{fig:gemini_imaging}}
\end{figure}

\textbf{Gemini-North:} TOI-4465 was observed on UT 2021 October 19 using the ‘Alopeke speckle instrument on the Gemini North 8-m telescope \citep{scott2021twin}. ‘Alopeke provides simultaneous speckle imaging in two bands (562~nm and 832~nm), with output data products including a reconstructed image with robust contrast limits on companion detections. Five sets of 1000~X~0.06 second images were obtained and processed with our standard reduction pipeline \citep{howell2011speckle}. The final contrast curves, and 832~nm reconstructed speckle image are shown in Figure \ref{fig:gemini_imaging}. We find that TOI-4465 is a single star with no companion brighter than 5-9.5 magnitudes below that of the target star from the 8-m telescope diffraction limit (20~mas) out to 1.2”. At the distance of TOI-4465 ($d \sim 122$~pc), these angular limits correspond to spatial limits of 2.4 to 146~AU.

\begin{figure}
    \centering
    \includegraphics[width=\linewidth]{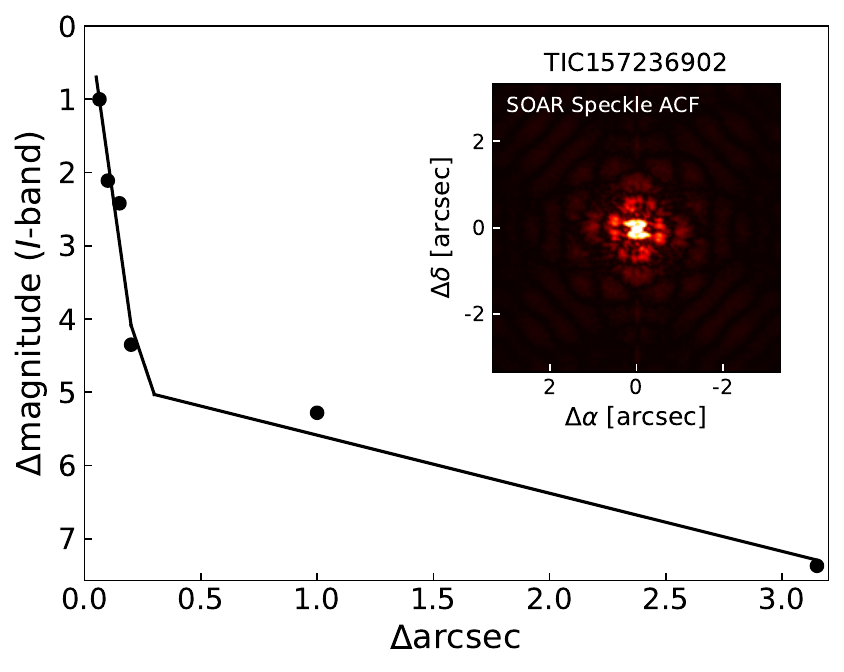}
\caption{SOAR speckle-imaging observations of TOI-4465. The contrast curve (5$\sigma$ sensitivity limits), and auto-correlation functions (inset) are shown. No additional sources were detected within 3\arcsec of TOI-4465. \label{fig:soar_imaging}}
\end{figure}

\textbf{SOAR:} We searched for stellar companions to TOI-4465 with speckle imaging on the 4.1-m Southern Astrophysical Research (SOAR) telescope \citep{tokovinin2018} on UT 2022 April 15, observing in Cousins I-band, a similar visible bandpass as TESS. This observation was sensitive to a 5.2-magnitude fainter star at an angular distance of 1\arcsec from the target. More details of the observations within the SOAR TESS survey are available in \citet{ziegler2020}. The 5$\sigma$ detection sensitivity and speckle auto-correlation functions from the observations are shown in Figure~\ref{fig:soar_imaging}. No nearby stars were detected within 3\arcsec of TOI-4465 in the SOAR observations. At the distance of TOI-4465 ($d \sim 122$~pc), these angular limits correspond to spatial limits of 6 to 366~AU.

\begin{figure}
    \centering
    \includegraphics[width=\linewidth]{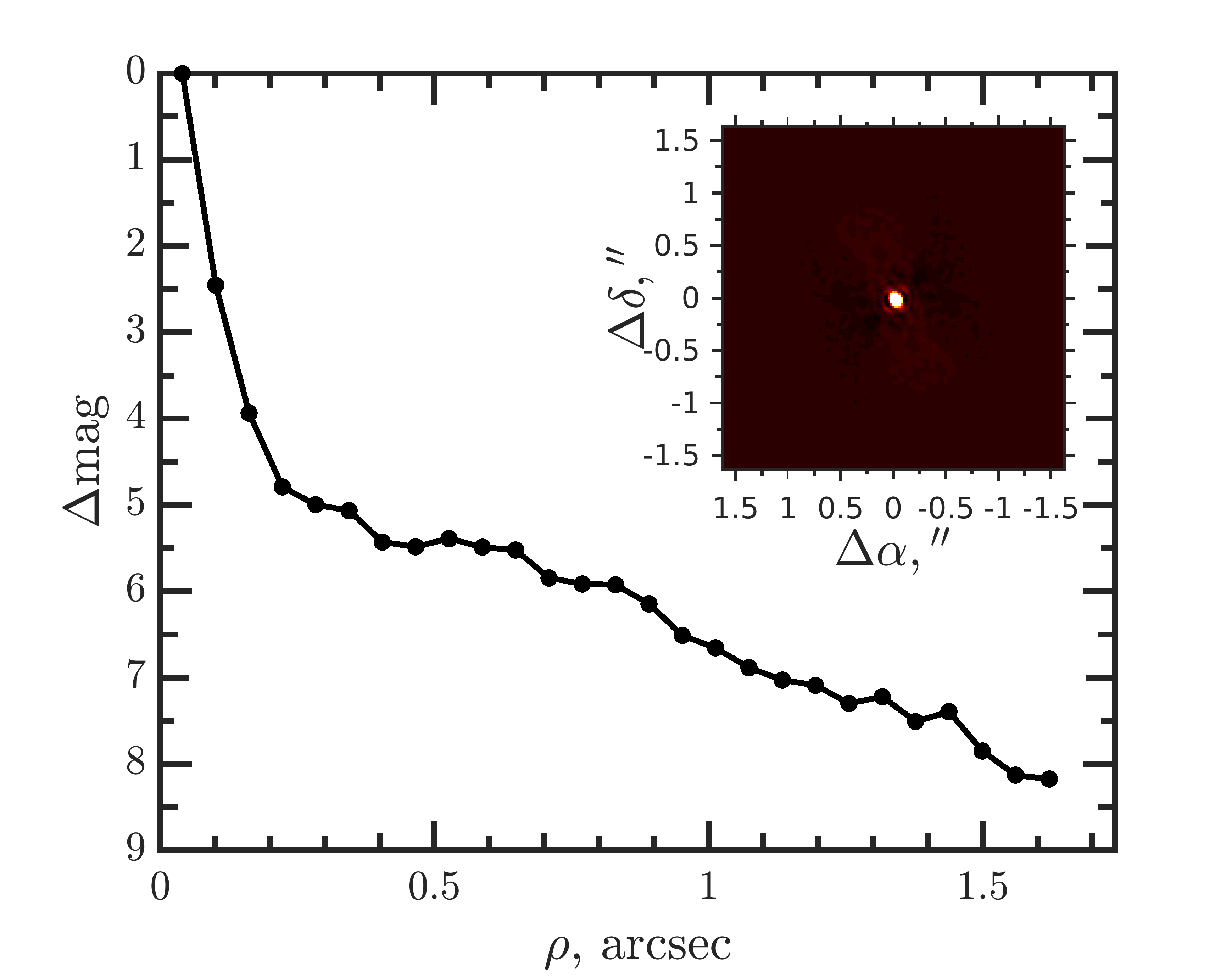}
\caption{SAI speckle-imaging observations of TOI-4465. The contrast curve (5$\sigma$ sensitivity limits), and auto-correlation functions (inset) are shown. \label{fig:SAI_imaging}}
\end{figure}

\textbf{SAI:} We observed TOI-4465 on UT 2021 October 22 with the Speckle Polarimeter \citep{Safonov2017} on the 2.5~m telescope at the Caucasian Observatory of Sternberg Astronomical Institute (SAI) of Lomonosov Moscow State University. Electron Multiplying CCD Andor iXon 897 was used as a detector. The atmospheric dispersion compensator allowed observation through the wide-band $I_c$ filter. The power spectrum was estimated from 4000 frames with 30~ms exposure. The detector has a pixel scale of $20.6$~mas/pixel, and the angular resolution was 89~mas. Long-exposure seeing was $0.92\arcsec$. We did not detect any stellar companions brighter than $\Delta I_C=5.0$ and $6.5$ at $\rho=0.25\arcsec$ and $1\arcsec$, respectively, where $\rho$ is the separation between the source and the potential companion (Figure~\ref{fig:SAI_imaging}). At the distance of TOI-4465 ($d \sim 122$~pc), these angular limits correspond to spatial limits of 11 to 122~AU.

\subsubsection{Near-Infrared AO Imaging}

\begin{figure}
    \centering
    \includegraphics[width=\linewidth]{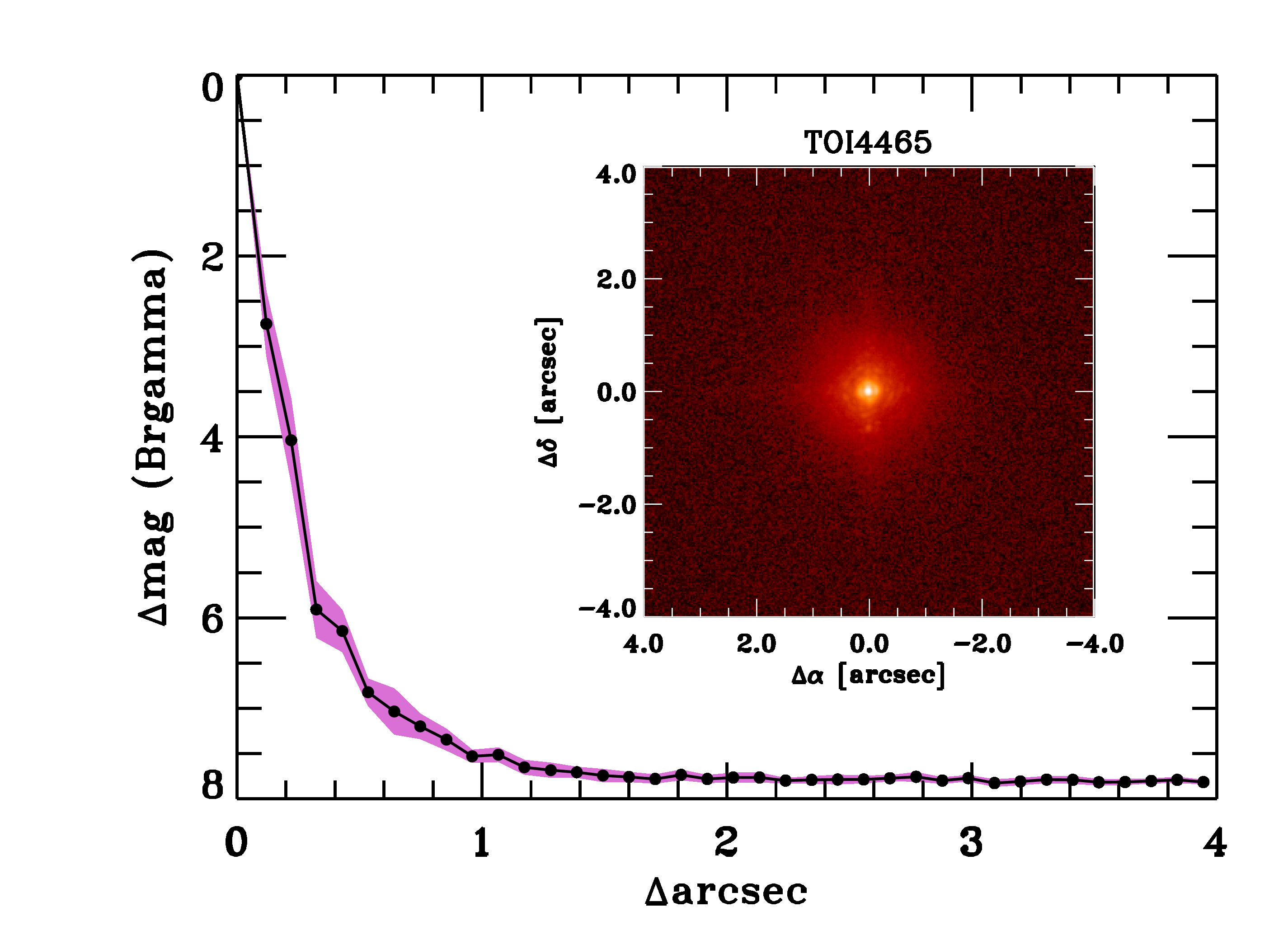}
\caption{NIR AO imaging and sensitivity curves for the Palomar Observations. {\it Insets:} Image of the central portion of the image.\label{fig:palomar_aoimaging}}
\end{figure}

Observations of TOI-4465 were made on 2021-Sep-20 with the PHARO instrument \citep{hayward2001} on the Palomar Hale (5 m) telescope behind the P3K natural guide star AO system \citep{dekany2013} in the narrowband Br-$\gamma$ filter $(\lambda_o = 2.1686; \Delta\lambda = 0.0326~\mu$m). The PHARO pixel scale is $0.025\arcsec$ per pixel. A standard 5-point quincunx dither pattern with steps of 5\arcsec\ was repeated twice with each repeat separated by 0.5\arcsec. The reduced science frames were combined into a single mosaiced image with a final resolution of 0.107\arcsec. The sensitivity of the final combined AO image was determined by injecting simulated sources azimuthally around the primary target every $20^\circ $ at separations of integer multiples of the central source's FWHM \citep{furlan2017}. The brightness of each injected source was scaled until standard aperture photometry detected it with $5\sigma $ significance.  The final $5\sigma $ limit at each separation was determined from the average of all of the determined limits at that separation and the uncertainty on the limit was set by the rms dispersion of the azimuthal slices at a given radial distance.  The Palomar sensitivities are shown in (Figure~\ref{fig:palomar_aoimaging}).  

\subsection{Parallax \& Reddening Estimates} 

Gaia DR3 reports TOI-4465 to have a parallax of $\varpi$ = $8.1756\pm0.0197$ mas \citep{GaiaDR3}, with corresponding geometric distance $d$ = $121.931^{+0.287}_{-0.285}$ pc \citep{BailerJones2021}, and modest proper motion of $\sim$70 mas\,yr$^{-1}$.

Reddening and extinction towards TOI-4465 appears to be very light, with previously published estimates of E(B-V) = $0.0108\pm0.0053$ \citep{Stassun2019}, E(B-V) = $0.0334\pm0.019$ \citep{Lallement2018}\footnote{STILISM reddening map at position of TOI-4665 for distance 122\,pc \citep{Lallement2018}. For negligibly reddened, late G-type dwarf stars (\Teff\, $\simeq$ 5500\,K) like \thisstar, $A_V$ $\simeq$ 3.2\,E(B-V) \citep[Fig. 4 of ][]{McCall2004}.}, and $A_0$ = $0.0016^{+0.0023}_{-0.0012}$ \citep{GaiaDR3}\footnote{Where $A_0$ $\simeq$ $A_V$.}

\subsection{Metallicity, Chemical Abundances, and Galactic Population}\label{sec:metallicity} 

We measured stellar atmospheric parameters from the high-resolution spectra obtained as part of our follow-up campaign. We used the \texttt{SpecMatch-Emp} code \citep{Yee2017_SpecMatchEmp}, which derives spectroscopic properties for a target star by matching its spectrum to a library of observed stellar spectra from stars with empirically-determined properties. We applied the code separately to high signal-to-noise template spectra from CHIRON and CORALIE, assembled by stacking the individual observations from each instrument. The results from the two analyses agree within 1-$\sigma$: we find $T_\mathrm{eff} = 5460 \pm 110$~K, $R_\star = 1.17 \pm 0.15\,R_\odot$, $\mathrm{[Fe/H]} = -0.19 \pm 0.09$~dex from the CHIRON observations; and $T_\mathrm{eff} = 5500 \pm 110$~K, $R_\star = 1.17 \pm 0.15\,R_\odot$, $\mathrm{[Fe/H]} = -0.09 \pm 0.09$~dex from the CORALIE data. The spectroscopic analyses indicate that TOI-4465 is a sub-solar-metallicity star.

Previously published estimates of [Fe/H], [M/H], and [$\alpha$/Fe] are listed in Table \ref{tab:metal}. Our measured metallicity, and the previously published [Fe/H] values are all consistent with \thisstar\, being slightly sub-solar by $\sim$0.1 dex. 
\citet{GaiaDR3} finds the star to be slightly enhanced in $\alpha$ elements by $\sim$0.1 dex.
However, the element-by-element chemical abundances for the $\alpha$ elements from the LAMOST survey show a somewhat unusual pattern.
The $\alpha$ elements O, Si, S, Ca, and Ti generally show abundances near solar or slightly enhanced ($\sim$0.1 dex) compared to the Sun,
however both Gaia DR3 and \citet{Xiang2019} find significantly super-solar [Mg/Fe] values: [Mg/Fe] = $0.205\pm0.017$ dex \citep{Xiang2019}, and [Mg/Fe] = $0.35^{+0.16}_{-0.13}$ \citep{GaiaDR3}. 
Overall the abundance pattern is consistent with \thisstar\, being a slightly sub-solar [Fe/H], approximately solar/slightly super-solar 
[$\alpha$/Fe], thin disk star, with the possible exception of enhanced Mg.

For the subsequent analyses in this paper, we adopt the stellar effective temperature and metallicity from LAMOST DR7 \citep{Luo2022}, namely $T_\mathrm{eff} = 5545 \pm 19$~K, and $\mathrm{[Fe/H]} = -0.06 \pm 0.016$~dex, since these values are more tightly constrained than those measured from the CHIRON/CORALIE spectra presented here.

Using the 3D space motion calculated with Gaia DR3 astrometry from \citet{GCNS2021} (barycentric $U, V, W$ = $-47.6, -32.8, 27.1$ \kms) with the Galactic kinematic membership formalism from \citet{Bensby2014}, we estimate Galactic kinematic membership probabilities of $P_{thin}$ = 93.1\%, $P_{thick}$ = 6.6\%, $P_{halo}$ = 0.002\%, and $P_{Hercules}$ = 0.33\%. 
From our chemical abundance analysis, consideration of previously published chemical abundances, and this kinematic membership calculation, we surmise that \thisstar\, is a Galactic thin disk star.

\begin{deluxetable}{ccc}[htb]
\tablewidth{\textwidth}
\tablecaption{Abundance Estimates for \thisstar \label{tab:metal}}
\tablehead{\colhead{Parameter} & \colhead{Value} & \colhead{Ref.}}
\startdata
\feh & $-0.03^{+0.14}_{-0.10}$ & 1$^a$\\
\feh & $-0.428\pm0.005$ & 1$^b$\\
\feh & $-0.060\pm0.016$ & 2\\
\feh & $-0.075\pm0.017$ & 3\\
\feh & $-0.19\pm0.09$ & 4\\
\mh  & $-0.09^{+0.02}_{-0.03}$ & 1$^a$\\
\mh  & $-0.37\pm0.05$ & 1$^c$\\
\afe & $0.07^{+0.03}_{-0.02}$ & 1$^a$\\
\afe & $0.11^{+0.03}_{-0.02}$ & 1$^c$\\
\cfe & $-0.031\pm0.019$ & 5\\
\nfe & $-0.020\pm0.036$ & 5\\
\ofe & $0.123\pm0.064$ & 5\\
\nafe & $-0.072\pm0.065$ & 5\\
\mgfe & $0.205\pm0.017$ & 5\\
\mgfe & $0.35^{+0.16}_{-0.13}$ & 1$^a$\\
\alfe & $0.134\pm0.032$ & 5\\
\sife & $-0.027\pm0.030$ & 5\\
\sife & $0.01\pm0.12$ & 1$^a$\\
\sfe  & $-0.03^{+0.17}_{-0.19}$ & 1$^a$\\
\cafe & $0.052\pm0.023$ & 5\\
\cafe & $0.08^{+0.04}_{-0.03}$ & 1$^a$\\
\tife & $0.007\pm0.023$ & 5\\
\tife & $0.19^{+0.21}_{-0.13}$ & 1$^a$\\
\crfe & $-0.068\pm0.025$ & 5\\
\mnfe & $-0.065\pm0.025$ & 5\\
\cofe & $-0.137\pm0.031$ & 5\\
\nife & $0.042\pm0.018$ & 5\\
\cufe & $0.073\pm0.187$ & 5\\
\bafe & $-0.180\pm0.117$ & 5\\
\enddata
\tablecomments{
(1) \citet{GaiaDR3},
(2) \citet{Luo2022},
(3) \citet{Luo2019},
(4) Section 2.4 (this work),
(5) \citet{Xiang2019}.
(a) from GSP-Spec MatisseGauguin using RVS spectra,
(b) GSP-Phot Aeneas best library using BP/RP spectra, 
(c) global metallicities from GSP-Spec ANN using RVS spectra.}
\end{deluxetable}

\section{Exoplanet Detection \& Follow Up \label{sec:planetdata}}

TOI-4465~b was first discovered as a single-transit event in TESS photometry (Section \ref{subsec:tessdata}). The TESS Follow-up Observing Program (TFOP)\footnote{\href{https://tess.mit.edu/followup}{https://tess.mit.edu/followup}} conducted numerous observations of TOI-4465.01/TOI-4465~b, including ground-based spectroscopic observations to measure the physical and orbital parameters of the planet (Section \ref{subsec:rvs}), and a global ground-based photometric campaign to observe another planetary transit (Section \ref{subsec:groundphotom}). We list all the contributed observations in the subsections below for completeness, however not every dataset is included in the final analysis. 

\subsection{\TESS\ Photometry \label{subsec:tessdata}}

TESS observed TOI-4465 (TIC~157236902) in Sector~26 (UT 2020 June 9 to 2020 July 4) of its primary mission, Sectors 40 (UT 2021 June 24 to 2021 July 23) and 53 (UT 2022 June 13 to 2022 July 9) of its first extended mission, and Sector 80 (UT 2024 June 18 to 2024 July 15) of its second extended mission. TOI-4465 was observed with 2-minute cadence in Sector~26, 10-minute cadence in Sectors 40 and 53, and 200-second cadence in Sector 80. The star fell on Camera~1 in Sectors 26, 40 and 53, and on Camera~2 in Sector 80. 

A single-transit event was identified in TESS Sector~40 by the Quick-Look Pipeline (QLP; \citealp{huang2020qlp,kunimoto2021qlp,kunimoto2022qlp}), which extracts light curves, and searches for transiting planet candidates from the TESS full-frame images (FFIs). The single-transit event was alerted as TOI-4465.01 by the \TESS\ TOI team on September 9, 2021 \citep{guerrero2021tess}. The TESS FFIs were calibrated by the Science Processing Operations Center (SPOC; \citealp{Jenkins2016}). The SPOC pipeline also identified the single-transit in the TESS-SPOC FFI light curve from Sector 40 \citep{TESSSPOC2020}, and the difference image centroid test \citep{Twicken:DVdiagnostics2018} located the host star to within $1.1 \pm 2.5$ of the transit source.

\begin{figure*}
    \centering
    \includegraphics[width=\textwidth,keepaspectratio]{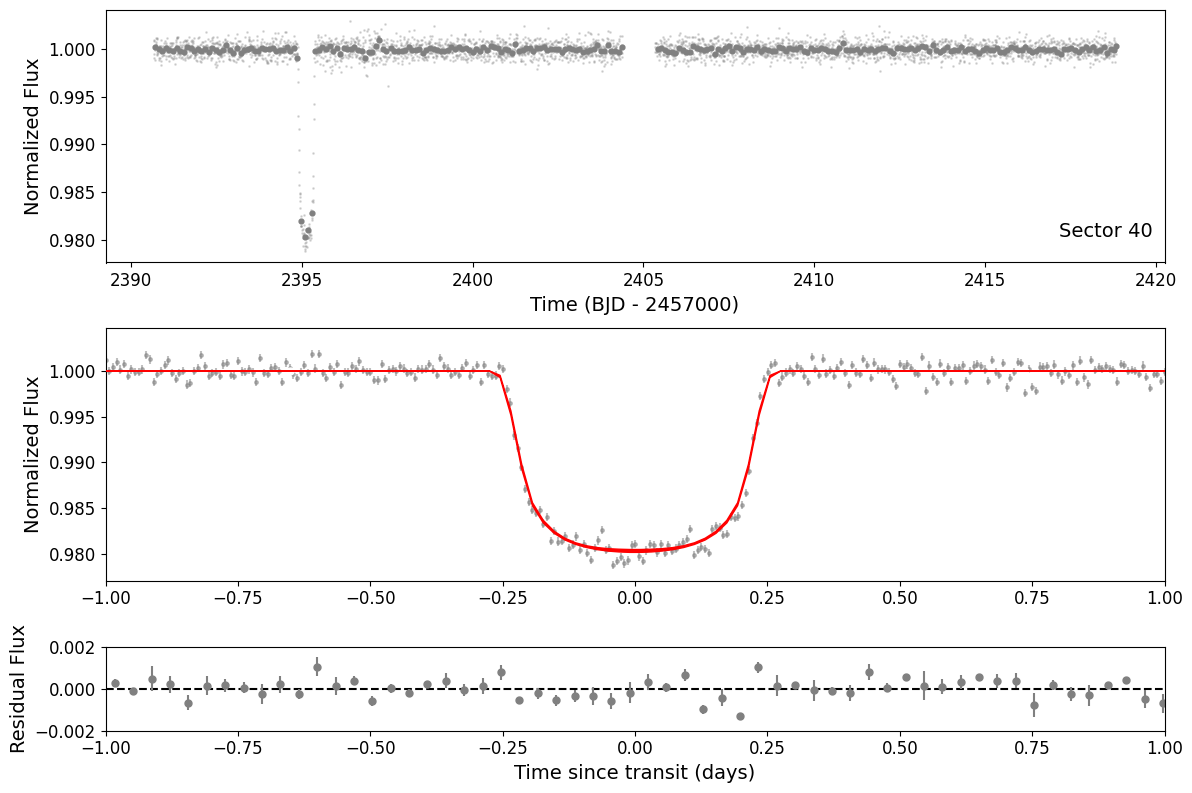}
\caption{\TESS\ light curves of \thisstar. \textit{Top:} Detrended, normalized and flattened Sector~40 FFI light curve. Lighter gray points are the \TESS\ 10-minute cadence flux measurements; darker points are the same data binned into 2.5-hour intervals. The single-transit occurred at 2459395.1256~BJD. \textit{Middle}: Zoomed-in view of the transit event, along with the best-fit transit model (red). Gray points are 10-minute cadence measurements. \textit{Bottom:} Residuals after the data have been subtracted from the best-fit model. Residuals are binned into 50-minute intervals. \label{fig:tessdataphased}}
\end{figure*}

We downloaded the Sector 40 FFI data from the Mikulski Archive for Space Telescopes (MAST)\footnote{\href{https://mast.stsci.edu}{https://mast.stsci.edu}} using the  \texttt{Lightkurve} package \citep{2018lightkurve}. Using \texttt{Lightkurve}, we extracted the light curve for TOI-4465 by selecting a 10x10 pixel region around the star and defining an aperture using a threshold-based mask centered on the target pixel with a flux threshold of 15 times the median pixel intensity of the image. Pixels outside the aperture were used to estimate the sky background. 
We modeled scattered light by constructing a design matrix from the background pixels, and applying principal component analysis (PCA) to capture the dominant scattered light trends. Linear regression against this background model was then used to subtract scattered light from the light curve.
We further detrended the FFI light curve by flattening the data using a Savitzky-Golay filter with a window size of $\sim5\%$ of the full length of the light curve, chosen to remove low-frequency noise while preserving transit shape, and by applying a median normalization to the light curve. The detrended \TESS\ Sector~40 FFI light curve, as well as a zoomed-in view of the transit event, are shown in Figure \ref{fig:tessdataphased}.

\subsection{Time Series Radial Velocities \label{subsec:rvs}}

Following the detection of the TESS single-transit event of TOI-4465.01, we began a spectroscopic monitoring campaign, measuring precise radial velocities (RVs) to confirm the planetary nature of the transit signal. The RV observations included in the final analysis are shown in Figure \ref{fig:RVsjointfit}.

\subsubsection{APF Observations} 

In September 2021, we began to acquire spectroscopy of TOI-4465 with the Levy spectrograph at the 2.4~m Automated Planet Finder (APF) telescope at Lick Observatory in California. The Levy spectrograph is a high-resolution ($R\sim114\,000$) slit-fed optical echelle spectrometer \citep{radovan2010radial,radovan2014automated,vogt2014apf}. This facility has a proven track record of acquiring precise Doppler spectroscopy in support of characterizing giant exoplanets with uncertain transit ephemerides \citep[e.g.,][]{dalba2022tess,mann2023giant,peluso2024confirming}.

We first acquired a reconnaissance spectrum with moderate S/N ($\sim$40) to determine if the host star was suitable for precise RVs. TOI-4465 showed negligible rotational velocity, and its spectrum did not contain a second set of lines. As a result, we proceeded with a nearly two-year long campaign that resulted in 56 RV measurements (Table~\ref{tab:APFrv}), which solved the orbital period of TOI-4465~b. The exposure times had a median value of 1800~s,which produced internal uncertainties of 4--10~m~s$^{-1}$. Within the spectrometer, light passes through a heated iodine cell that allows for wavelength calibration and the forward modeling of the stellar RV \citep{butler1996attaining,fulton2015kelt}. This procedure requires a high-S/N spectrum to be used as a template. For this, we employed the High Resolution Echelle Spectrometer \citep[HIRES;][]{vogt2014apf} on the Keck~I telescope. The template spectrum was obtained on UT 2022 February 11, and has a S/N value of $\sim$200.

\begin{center}
\begin{longtable}{ccc}
\caption{APF RV data of \thisstar}\\
\hline Date (BJD$_{\rm TDB}$) & RV ($\ms$) & $\sigma_{RV}$ ($\ms$)\\
\hline \vspace{2pt}
2459468.77560    & 324.74            & 6.97              \\ 
2459470.79754    & 317.99            & 6.30              \\ 
2459475.74011    & 292.13            & 6.47              \\ 
2459486.75338    & 149.51            & 17.09             \\ 
2459489.75960    & 135.32            & 6.26              \\ 
2459498.69482    & 53.29             & 5.95              \\ 
2459516.59819    & -149.06           & 4.91              \\ 
2459522.62827    & -167.15           & 6.81              \\ 
2459532.64852    & -204.25           & 9.18              \\ 
2459543.57855    & -162.42           & 6.60              \\ 
2459546.59434    & -60.84            & 8.65              \\ 
2459550.57934    & 53.00             & 5.16              \\ 
2459560.57723    & 280.03            & 9.55              \\ 
2459591.09198    & 150.04            & 8.16              \\ 
2459598.08032    & 74.30             & 7.11              \\ 
2459603.06851    & -7.38             & 6.47              \\ 
2459605.07575    & -6.18             & 7.30              \\ 
2459607.08546    & -28.90            & 7.59              \\ 
2459614.06288    & -89.53            & 9.26              \\ 
2459619.06725    & -141.77           & 6.81              \\ 
2459625.00123    & -172.16           & 6.85              \\ 
2459630.00425    & -217.35           & 6.95              \\ 
2459639.96430    & -198.98           & 6.95              \\ 
2459645.97296    & -102.05           & 10.41             \\ 
2459650.97696    & 12.37             & 6.15              \\ 
2459656.93428    & 168.07            & 7.34              \\ 
2459661.94932    & 278.24            & 6.13              \\ 
2459671.97022    & 320.08            & 6.68              \\ 
2459701.99856    & 57.06             & 8.64              \\ 
2459717.87669    & -146.73           & 7.59              \\ 
2459732.88613    & -204.10           & 5.26              \\ 
2459754.99205    & 76.24             & 5.77              \\ 
2459755.97648    & 97.46             & 5.89              \\ 
2459769.95910    & 348.52            & 5.76              \\ 
2459783.91919    & 264.77            & 6.33              \\ 
2459802.74549    & 52.60             & 4.71              \\ 
2459817.77572    & -110.71           & 4.90              \\ 
2459817.77572    & -110.71           & 4.90              \\ 
2459835.70532    & -209.96           & 5.42              \\ 
2459835.70532    & -209.96           & 5.42              \\ 
2459850.70414    & -84.79            & 4.83              \\ 
2459850.70414    & -84.79            & 4.83              \\ 
2459868.62918    & 324.99            & 4.63              \\ 
2459895.58760    & 154.68            & 5.28              \\ 
2459914.57983    & -56.32            & 6.40              \\ 
2459966.08045    & 246.61            & 8.28              \\ 
2459993.04966    & 202.40            & 5.68              \\ 
2460043.98914    & -226.15           & 4.82              \\ 
2460121.76455    & -86.11            & 4.73              \\ 
2460124.97573    & -122.59           & 6.45              \\ 
2460128.90406    & -148.02           & 5.54              \\ 
2460132.87999    & -163.18           & 5.18              \\ 
2460135.98455    & -195.34           & 5.64              \\ 
2460139.87669    & -217.53           & 4.78              \\ 
2460143.83039    & -224.53           & 5.54              \\ 
2460150.78727    & -181.64           & 4.43              \\ 
\hline
\label{tab:APFrv}
\end{longtable}
\end{center}

\subsubsection{CHIRON Observations} 

We observed TOI-4465 with CHIRON \citep{chiron} on 15 different nights, between UT 2022 July 31 and 2023 June 11, as part of a RV follow-up program, within the Warm gIaNts with TESS (WINE) collaboration \citep{brahm2019, jones2024}. 
CHIRON is a high-resolution fiber-fed spectrograph mounted on the 1.5\,m telescope at the Cerro Tololo Inter-American Observatory (CTIO) in Chile.
For the observations we adopted an exposure time of 1200~s, leading to a S/N per extracted pixel between $\sim$ 20-30, depending on the airmass and the atmospheric conditions. In addition, we used the image-slicer (R $\sim$ 80,000), and we corrected the spectral drift by taking a ThAr lamp spectrum immediately after the science spectrum. The echelle raw data were extracted and calibrated with the CHIRON pipeline \citep{Paredes2021}, and the RV measurements (Table \ref{tab:CHIRONrv}) were computed with the  pipeline presented in \citet{Jones2019}.

\begin{center}
\begin{longtable}{ccc}
\caption{CHIRON RV data of \thisstar}\\
\hline Date (BJD$_{\rm TDB}$) & RV ($\ms$) & $\sigma_{RV}$ ($\ms$)\\
\hline \vspace{2pt}
2459791.65808 & -11640.1         & 10.0            \\
2459820.54855 & -11945.7         & 8.7             \\
2459827.53272 & -11989.4         & 6.3             \\
2459828.49990 & -12006.0         & 9.4             \\
2460036.90349 & -12052.8         & 10.0            \\
2460043.85943 & -12061.5         & 8.0             \\
2460057.88370 & -11851.6         & 11.3            \\
2460062.84909 & -11695.1         & 9.5             \\
2460066.86861 & -11603.8         & 7.1             \\
2460071.81724 & -11541.7         & 11.9            \\
2460077.81186 & -11490.1         & 12.5            \\
2460085.79166 & -11530.2         & 7.7             \\
2460096.82560 & -11660.5         & 9.0             \\
2460101.79301 & -11688.0         & 8.6             \\
2460106.75398 & -11743.1         & 11.0            \\
\hline
\label{tab:CHIRONrv}
\end{longtable}
\end{center}

\subsubsection{TRES Observations} 

We used the Tillinghast Reflector Echelle Spectrograph (TRES; \citealp{Furesz2008}) mounted on the 1.5~m Tillinghast Reflector telescope at the Fred Lawrence Whipple Observatory (FLWO) atop Mount Hopkins, Arizona to obtain 14 spectra of TOI-4465 between UT 2021 September 11 and 2022 July 11. 
TRES is an optical, fiber-fed echelle spectrograph with a wavelength range of 390-910~nm, and a resolving power of R$\sim$44,000. The exposure time varied from 220-720 seconds depending on observing conditions, and the average S/N per resolution element was $\sim$37.
The spectra were extracted as described in \citet{Buchhave2010hatp16}. A multi-order analysis was then performed by cross-correlating the strongest observed spectrum as a template, order by order, against the remaining spectra. The spectra were then used to derive stellar parameters using the Stellar Parameter Classification tool (SPC; \citealp{Buchhave2012}). SPC cross correlates an observed spectrum against a grid of synthetic spectra based on the Kurucz atmospheric model \citep{Kurucz1992}. 
The derived spectral parameters were: $T_\mathrm{eff} = 5599 \pm 50$~K,  $\logg = 4.44 \pm 0.10$ and $[m/H] = -0.05 \pm 0.08$, which are broadly consistent with values derived in Section \ref{sec:stellardata}.

\begin{center}
\begin{longtable}{ccc}
\caption{TRES RV data of \thisstar}\\
\hline Date (BJD$_{\rm TDB}$) & RV ($\ms$) & $\sigma_{RV}$ ($\ms$)\\
\hline \vspace{2pt}
2459468.65264	&	558.89	&	19.55	\\
2459478.60722	&	522.93	&	23.28	\\
2459489.64825	&	436.15	&	25.26	\\
2459502.59834	&	362.80	&	23.83	\\
2459703.85223	&	253.15	&	19.30	\\
2459708.98061	&	226.38	&	25.55	\\
2459715.86088	&	128.21	&	21.45	\\
2459722.88156	&	92.33	&	24.60	\\
2459732.93949	&	0.00	&	23.83	\\
2459740.97807	&	26.59	&	17.47	\\
2459743.96380	&	76.52	&	29.40	\\
2459750.85263	&	215.69	&	19.77	\\
2459766.79061	&	562.64	&	27.60	\\
2459771.86215	&	546.57	&	26.20	\\
\hline
\label{tab:TRESrv}
\end{longtable}
\end{center}

\begin{figure*}
    \centering
    \includegraphics[width=\textwidth,keepaspectratio]{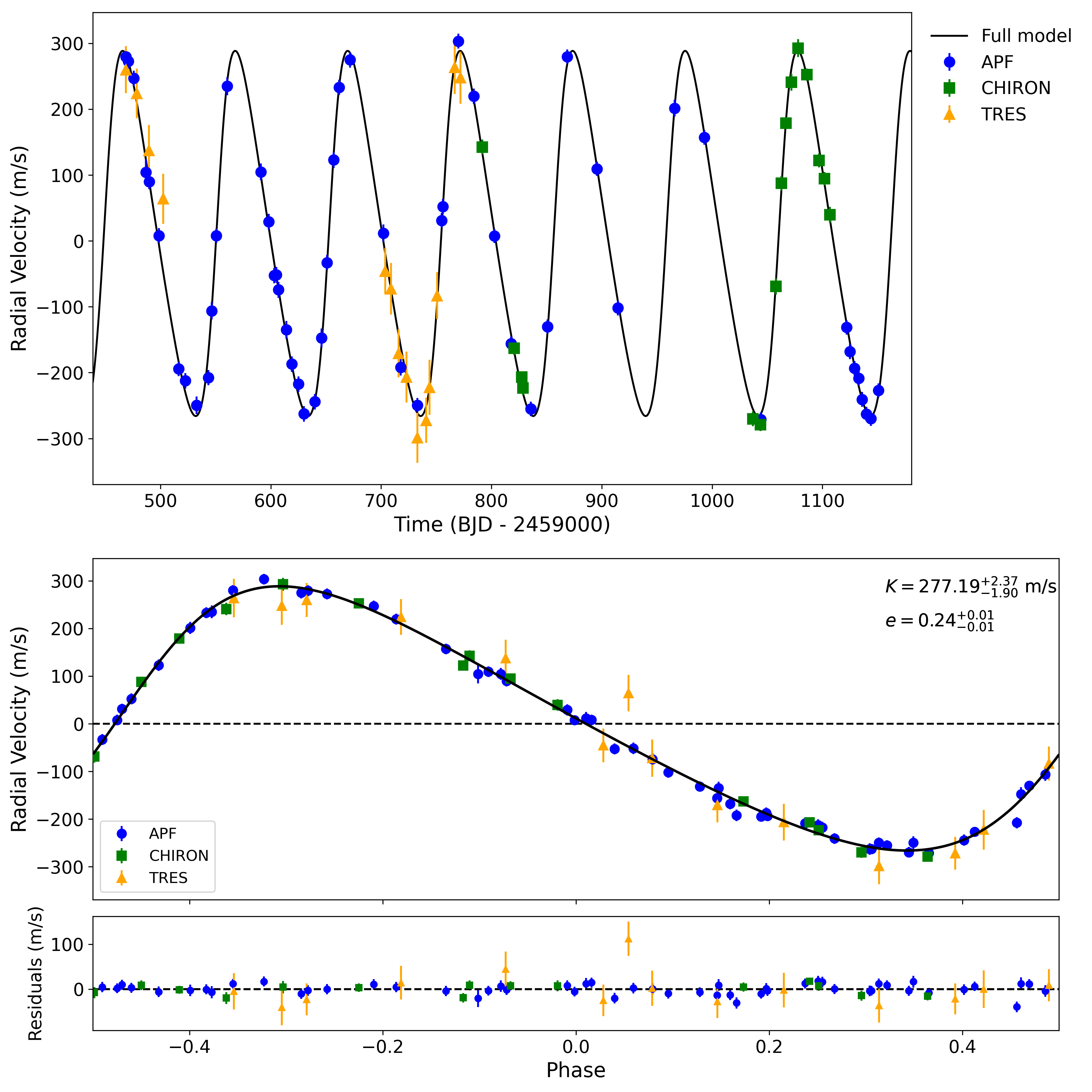}
\caption{Results of the \texttt{juliet} joint fit to the APF, CHIRON and TRES radial velocities. \textit{Top:} APF (blue points), CHIRON (green points) and TRES (orange points) RVs over time, and full best-fit \texttt{juliet} model (black curve). \textit{Middle:} Phase-folded RV measurements from APF, CHIRON and TRES. The black curve is the best-fit \texttt{juliet} RV model. Errorbars are the quadrature sum of the indvidual instrument internal uncertainties, and the RV jitter estimate from the \texttt{juliet} fit. The best-fit RV semi-amplitude and orbital eccentricity are $K = 277.19^{+2.37}_{-1.90}~\ms$ and $e=0.24\pm0.01$, respectively. \textit{Bottom:} RV residuals after the data have been subtracted from the best-fit model. 
\label{fig:RVsjointfit}}
\end{figure*}

\subsubsection{CORALIE Observations \label{subsubsec:coralie}} 

TOI-4465 was monitored with the high-resolution spectrograph CORALIE \citep{queloz2001coralie} for one month from UT 2022 April 7 to 2022 May 7. CORALIE is a fiber-fed echelle spectrograph installed at the Nasmyth focus of the 1.2~m Euler telescope at La Silla Observatory in Chile. CORALIE has a spectral resolution of $R\sim60,000$ with a 3-pixel sampling per resolution element. We obtained six spectra of TOI-4465, each with an exposure time of 1800 seconds, resulting in a S/N value ranging from 20 and 30 at 550~nm, depending on observing conditions. The spectra were reduced with the standard data reduction pipeline using a G2 stellar mask to compute radial velocity measurements via the cross-correlation technique \citep[e.g.][]{pepe2002harps}. The radial velocities and associated errors from the CORALIE observations are reported in Table~\ref{tab:CORALIErv}. 

We estimate the stellar projected rotational velocity ($v\sin i$) using a calibration based on the full width at half maximum (FWHM) of the cross-correlation function, and the B-V magnitude of the star. The FWHM was determined to be $8.048\pm 0.011$~\kms\, from the CORALIE spectra, leading to an upper limit on $v\sin i$ of 4.13~\kms, as this value is at the calibration set's working limit for a FWHM of 8.1~\kms.

\begin{center}
\begin{longtable}{ccc}
\caption{CORALIE RV data of \thisstar}\\
\hline Date (BJD$_{\rm TDB}$) & RV ($\ms$) & $\sigma_{RV}$ ($\ms$)\\
\hline \vspace{2pt}
2459676.88327 & -49306.6         & 6.7             \\
2459686.90125 & -49417.4         & 6.5             \\
2459691.90006 & -49458.5         & 9.0             \\
2459700.88245 & -49552.7         & 13.9            \\
2459705.82078 & -49629.1         & 14.7            \\
2459706.83717 & -49642.2         & 8.8             \\
\hline
\label{tab:CORALIErv}
\end{longtable}
\end{center}

We did not include the CORALIE RVs in the final global model fit because the data increased parameter uncertainties, and added noise without enhancing model precision (see Section \ref{sec:juliet_jointfit}).

\begin{figure*}[htb!]
    \centering
    \includegraphics[width=\textwidth,keepaspectratio]{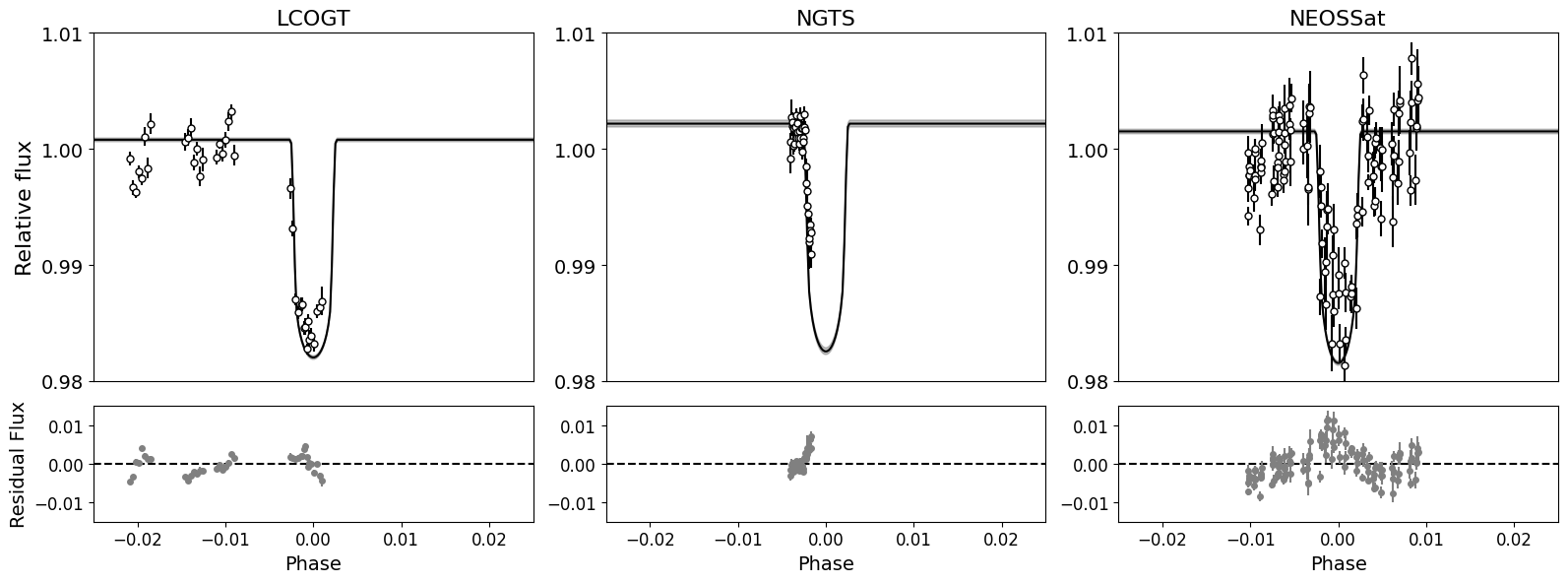}
\caption{Results of the \texttt{juliet} joint fit to LCOGT, NGTS and NEOSSat photometry. \textit{Top:} Transit observations of the respective instruments phase-folded to the period of \thisplanetb. The black curve is the best-fit \texttt{juliet} transit model, and the 68\% confidence interval is represented by the gray shaded region. Binned data points with errorbars are shown for clarity (white circles). The LCOGT data are binned in 45-minutes intervals, the NGTS data are binned in 10-minute intervals, and the NEOSSat data are binned in 4-minute intervals. \textit{Bottom:} Residuals after the data have been subtracted from the best-fit model.
\label{fig:transitsjointfit}}
\end{figure*}

\subsection{Ground-based Time-Series Photometry \label{subsec:groundphotom}} 

Based on an ephemeris that was predicted from the TESS single transit event and the precise radial velocities collected as part of this work, we conducted a photometric follow-up campaign as part of the \textit{TESS} Follow-up Observing Program Sub Group 1 \citep[TFOP SG1;][]{collins:2019} to determine the precise orbital period of the planet. We used the {\tt TESS Transit Finder}, which is a customized version of the {\tt Tapir} software package \citep{Jensen:2013}, to schedule our transit observations.

\subsubsection{General TFOP Follow-up}

Using the large TFOP SG1 network of observatories, we observed TOI-4465 from UT 2022 August 8 to 2022 August 13 (2459800.7478 - 2459804.8129 BJD$_{\rm TDB}$) with a goal of near continuous longitudinal coverage to minimize the number of gaps in the lightcurve data. The observatories that contributed light curve data to the campaign are detailed in Table \ref{tab:transitfollowup}, along with a summary of each result. 

Las Cumbres Observatory Global Telescope \citep[LCOGT;][]{Brown:2013} images were calibrated by the LCOGT {\tt BANZAI} pipeline \citep{McCully:2018}. Except for the NGTS and Unistellar observations described below, all differential photometric data were extracted using {\tt AstroImageJ} \citep{Collins:2017}.

We detected a $\sim$13 ppt ingress on-target near 2459802.63~BJD$_{\rm TDB}$ (Figure \ref{fig:transitsjointfit}). The light curve data are available on the {\tt EXOFOP-TESS} website\footnote{\href{https://exofop.ipac.caltech.edu/tess/target.php?id=157236902}{https://exofop.ipac.caltech.edu/tess/target.php?id=157236902}} and are included in the global modeling described in Section \ref{sec:juliet_jointfit}.

\subsubsection{NGTS Observations} 

TOI-4465 was observed by the Next Generation Transit Survey (NGTS; \citealp{wheatley2018next}) for six nights from UT 2022 August 8 to 2022 August 14, resulting in the observation of a partial transit on 2022 August 10 ($\sim$2459802.7~BJD$_{\rm TDB}$; see Figure \ref{fig:transitsjointfit}). 
NGTS is an array of twelve 20~cm-diameter telescopes at the Paranal site of the European Southern Observatory. Observations can be taken in blind-service mode with a single telescope or, in the case of important targets, observed with multiple cameras simultaneously. However, given the depth of the transit around TOI-4465, only a single NGTS camera was required for these observations. NGTS can achieve a photometric precision of 400~ppm in 30 minutes using a single camera for targets brighter than $G = 12$~mag.
All NGTS data were taken with an exposure time of 10 seconds, and a cadence of 13 seconds. Magnitude data was extracted from the telescope using the purpose-built Bright Stars Process pipeline described in \citet{bryant2020simultaneous}.\footnote{\href{https://github.com/EMBryant/bsproc}{https://github.com/EMBryant/bsproc}}
An aperture of 4-pixels was found to best preserve the transit signal while reducing the effects of additional systematics. The partial transit seen by NGTS was observed concurrently to the LCOGT observation, corroborating the transit timing.

\subsubsection{Unistellar Observations} 

We employed the Unistellar Citizen Science Network to observe TOI-4465 at various times from UT 2022 August 9 to 2022 August 13. The Unistellar Network is a global collaboration of citizen scientists using Unistellar telescopes called eVscopes, digital Newtonian reflectors (114~mm apertures, 450~mm focal length) equipped with SONY CMOS imaging sensors and a mobile application to aid in operation and data upload \citep{Marchis2020a}. These CMOS sensors are sensitive to blue, green, and red bandpasses via a Bayer filter. Once an observer has taken transit data, their observation is uploaded and run through a custom python pipeline for analysis. The structure of the Unistellar Network Exoplanet Program is described in detail in \citet{peluso2023unistellar}.

In total, 24 observers from the USA, UK, Austria, New Zealand, Germany, Italy, France, the Netherlands, Japan, and Switzerland participated in the TOI-4465 ground-based photometry campaign; the observations that were used in our final analysis of the TOI-4465 system (13 datasets), as well as observations that ruled out the transit, are detailed in Table \ref{tab:transitfollowup}. Observers gathered data with a 3.97s exposure time and cadence, and did not use any additional filters. For each eVscope dataset, images that were off-target or saturated were removed, and remaining images were dark subtracted (when dark frames were available) and plate-solved. These calibrated images were averaged into stacked images with a two-minute total integration time to increase S/N. We then perform differential aperture photometry using an optimized combination of up to 10 reference stars median combined to form a “composite reference star”, and an optimal aperture size gathered from the FWHM of the target star over all frames \citep{Dalba2016, Dalba2017a}. A detailed overview of the photometry methodology used for Unistellar data is given in \citet{sgro2024confirmation}. Note that we detrend for airmass only for fully out-of-transit datasets. 

\begin{figure*}
    \centering
    \includegraphics[width=\textwidth,keepaspectratio]{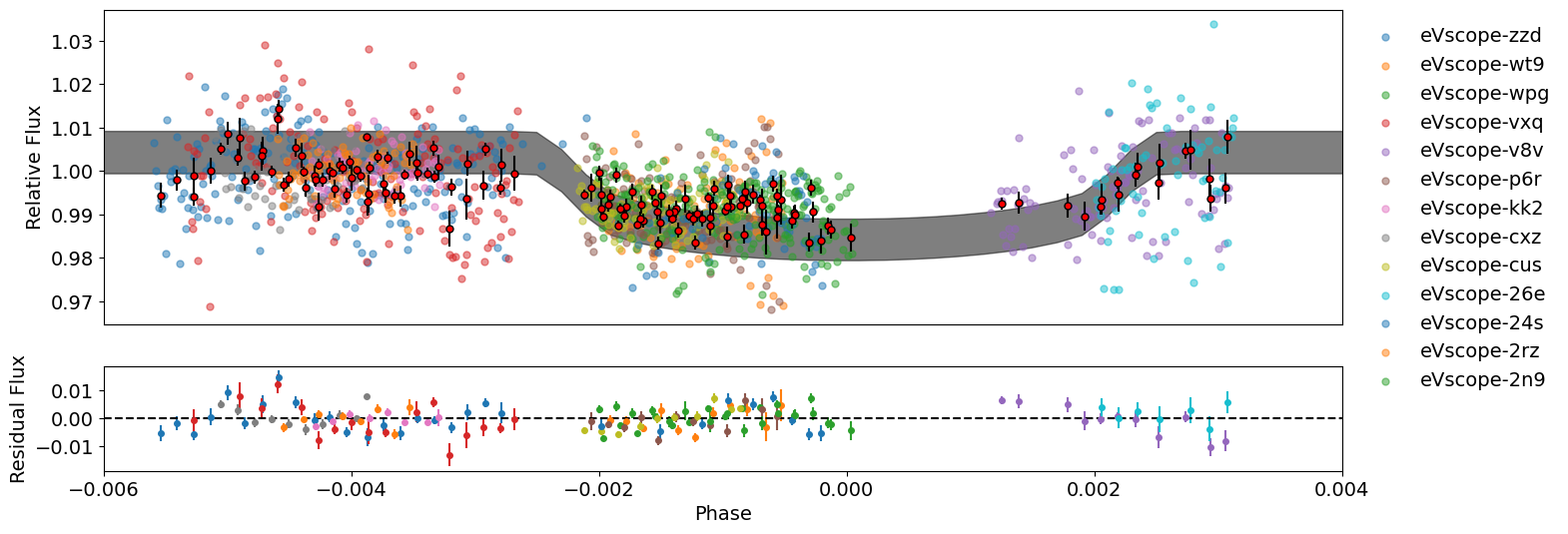}
\caption{Results of the \texttt{juliet} joint fit to the Unistellar photometry. \textit{Top:} Photometric data from the 13 eVscopes phase-folded to the period of \thisplanetb. The gray band is the best-fit \texttt{juliet} transit model with $1\sigma$ uncertainty. The binned data (red points) with errorbars are shown for clarity. The data are binned in 20-minute intervals. \textit{Bottom:} Residuals after the data have been subtracted from the best-fit model.
\label{fig:unistellartransitsjointfit}}
\end{figure*}

We detected an egress signal on target on UT 2022 August 11 ($\sim$2459803.15~BJD$_{\rm TDB}$). The composite light curve of Unistellar in-transit and out-of-transit data that cover the transit duration and relevant baseline are included in Figure~\ref{fig:unistellartransitsjointfit} and the global system modeling described in Section~\ref{sec:juliet_jointfit}.

\startlongtable
\begin{deluxetable*}{lclccl}
\tabletypesize{\small}
\tablewidth{\linewidth}
\tablecaption{Ground-based light curve observations for TOI-4465.01 \label{tab:transitfollowup} }
\tablehead{\colhead{Observatory} & \colhead{Aperture (m)} & \colhead{Location} & \colhead{Filter} & \colhead{Start Obs - End Obs$^{1}$} & \colhead{Result}}
\startdata
eVscope$^{2}$-2rz & 0.1 & Puy-Saint-Vincent, France & Clear$^{3}$ & 9800.4387 – 9800.5575 & ruled out \\
eVscope-rev & 0.1 & Austin, Texas, USA  & Clear & 9800.6128 – 9800.8671 & ruled out \\
eVscope-qer & 0.1 & Raleigh, North Carolina, USA & Clear & 9800.6137 – 9800.7415 & ruled out \\
eVscope-jjh & 0.1 & Longmont, Colorado, USA  & Clear & 9800.6710 – 9800.8518 & ruled out \\
eVscope-2n9 & 0.1 & Richmond, California, USA & Clear & 9800.6880 – 9800.9033 & ruled out \\
eVscope-wpg & 0.1 & Richmond, California, USA & Clear & 9800.6921 – 9800.9063 & ruled out \\
LCOGT$^{4}$-Hal & 0.4 & Maui, Hawai'i, USA & $z$-short &  9800.7478 -- 9801.0016 & ruled out \\
LCOGT-SSO & 0.4 & Coonabarabran, Australia & $z$-short$^{5}$ & 9800.8576 -- 9801.0518 & ruled out \\
Hazelwood Obs & 0.3 & Churchill, Victoria, Australia & R &  9800.8855 -- 9801.0727 & inconclusive \\
Grand-Pra Obs & 0.4 & Sion, Valais, Switzerland & Sloan $i'$ & 9801.3361 --  9801.4995 & ruled out \\
Kotizarovci Obs & 0.3 & Viskovo, Croatia & Baader R$^{6}$ & 9801.3381 -- 9801.5655 & ruled out \\
eVscope-zzd & 0.1 & Vienna, Austria & Clear & 9801.3169 – 9802.6036 & ruled out \\
eVscope-kk2 & 0.1 & Finowfurt, Brandenburg, Germany & Clear & 9801.3284 – 9801.4653 & ruled out \\
eVscope-knr & 0.1 & Bayonne, Aquitaine, France & Clear & 9801.3438 – 9801.5136 & ruled out \\
eVscope-gv6 & 0.1 & Valeggio sul Mincio, Verona, Italy & Clear & 9801.3571 – 9801.4438 & ruled out \\
eVscope-vxq & 0.1 & Saxon-Sion, France & Clear & 9801.3577 – 9801.6520 & ruled out \\
eVscope-hkf & 0.1 & Saint Paul les Dax, France & Clear & 9801.3920 – 9801.4335 & ruled out \\
OBP$^{7}$ & 0.4 & Moydans, France & R &  9801.4700 -- 9801.5633 & ruled out \\
eVscope-rev & 0.1 & Austin, Texas, USA & Clear & 9801.6000 – 9801.8548 & ruled out \\    
eVscope-jjh & 0.1 & Longmont, Colorado, USA & Clear & 9801.6217 – 9801.9112 & ruled out \\
eVscope-24s  & 0.1 & Ames, Iowa, USA & Clear & 9801.6576 – 9801.8326 & ruled out \\
eVscope-wt9 & 0.1 & San Leandro, California, USA & Clear & 9801.6643 – 9801.7607 & ruled out \\ 
eVscope-p6r & 0.1 & San Leandro, California, USA & Clear & 9801.6676 – 9801.7607 & ruled out \\
eVscope-wpg & 0.1 & Richmond, California, USA & Clear & 9801.6775 – 9801.8971 & ruled out \\
eVscope-2n9 & 0.1 & Richmond, California, USA & Clear & 9801.6812 – 9801.7982 & ruled out \\
eVscope-cus & 0.1 & Temecula, California, USA & Clear & 9801.7230 – 9801.8538 & ruled out \\
eVscope-t26 & 0.1 & Oakland, California, USA & Clear & 9801.7339 – 9801.8195 & ruled out \\
LCOGT-Hal & 0.4 & Maui, Hawai'i, USA & $z$-short &  9801.7472 -- 9802.0018 & ruled out \\
eVscope-4bj & 0.1 & Christchurch, New Zealand & Clear & 9801.7847 – 9801.9181 & ruled out \\
eVscope-e7m & 0.1 & Tonosyo, Kagawa, Japan & Clear & 9801.9916 – 9802.1198 & ruled out \\
eVscope-v8v & 0.1 & Ishioka, Ibaraki, Japan & Clear & 9802.0454 – 9802.0605 & ruled out \\
eVscope-zzd & 0.1 & Vienna, Austria & Clear & 9802.3126 – 9802.6061 & ruled out \\
Kotizarovci Obs & 0.3 & Viskovo, Croatia & T &  9802.3335 -- 9802.5678 & ruled out \\
OBP & 0.4 & Moydans, France & R &  9802.3340 -- 9802.5266 & ruled out \\
Grand-Pra Obs & 0.4 & Sion, Valais, Switzerland & Sloan $i'$ & 9802.3394 -- 9802.4115 & inconclusive \\
eVscope-vxq & 0.1 & Saxon-Sion, France & Clear & 9802.3403 – 9802.6274 & ruled out \\
eVscope-cxz & 0.1 & Tonbridge, England & Clear & 9802.3620 – 9802.4879 & ruled out \\
RFAC$^{8}$ & 0.3 & Fregenal de la Sierra, Spain & R & 9802.3813 -- 9802.5192 & ruled out \\
eVscope-2rz & 0.1 & Puy-Saint-Vincent, France & Clear & 9802.4140 – 9802.5275 & ruled out \\
eVscope-kk2 & 0.1 & Finowfurt, Brandenburg, Germany & Clear & 9802.4401 – 9802.5541 & ruled out \\
NGTS & 0.2 & Paranal Observatory, Chile & 520--890 nm &  9802.4712 -- 9802.7242 & ingress \\
eVscope-24s & 0.1 & Ames, Iowa, USA & Clear & 9802.6087 – 9802.8636 & ingress \\
LCOGT-McD & 0.4 & McDonald Obs, TX, USA & $z$-short & 9802.6152 -- 9802.8369 & ingress \\
MLO$^{9}$ & 0.4 & Glendora, CA, USA & I & 9802.6608 -- 9802.9152 & in transit \\
eVscope-cus & 0.1 & Temecula, California, USA & Clear & 9802.6611 – 9802.7958 & in transit \\
eVscope-p6r & 0.1 & San Leandro, California, USA & Clear & 9802.6666 – 9802.8314 & in transit \\
eVscope-2n9 & 0.1 & Richmond, California, USA & Clear & 9802.6737 – 9802.8881 & in transit \\
eVscope-wpg & 0.1 & Richmond, California, USA & Clear & 9802.6772– 9802.8894 & in transit \\
eVscope-wt9 & 0.1 & San Leandro, California, USA & Clear & 9802.6825 – 9802.8314 & in transit \\
LCOGT-Hal & 0.4 & Maui, Hawai'i, USA & $z$-short &  9802.7466 -- 9803.0010 & in transit \\
eVscope-v8v & 0.1 & Ishioka, Ibaraki, Japan & Clear & 9803.0016 – 9803.1987 & egress \\
eVscope-26e & 0.1 & Ishioka, Ibaraki, Japan & Clear & 9803.0842 – 9803.2023 & egress \\
eVscope-zzd & 0.1 & Vienna, Austria & Clear & 9803.3070 – 9803.5816 & ruled out \\
eVscope-kk2 & 0.1 & Finowfurt, Brandenburg, Germany & Clear & 9803.3201 – 9803.5027 & ruled out \\
eVscope-vfq & 0.1 & Lattes, France & Clear & 9803.3234 – 9803.5022 & ruled out \\
OBP & 0.4 & Moydans, France & R &  9803.3279 -- 9803.5120 & ruled out \\
eVscope-2rz & 0.1 & Puy-Saint-Vincent, France & Clear & 9803.3311 – 9803.5425 & ruled out \\
eVscope-vxq & 0.1 & Saxon-Sion, France & Clear & 9803.3417 – 9803.5393 & ruled out \\
eVscope-n47 & 0.1 & Laren gld, Netherlands & Clear & 9803.3523 – 9803.5217 & ruled out \\
RFAC & 0.3 & Fregenal de la Sierra, Spain & R & 9803.3752 -- 9803.5664 & ruled out \\
OAUV$^{10}$ & 0.5 & Valencia, Spain & R &  9803.3906 -- 9803.5553 & ruled out \\
OAA$^{11}$ & 0.4 & Albanya, Girona & I & 9803.4649 -- 9803.5792 & ruled out \\
eVscope-934 & 0.1 & Athens, Georgia, USA  & Clear & 9803.5517 – 9803.5933 & ruled out \\
eVscope-rev & 0.1 & Austin, Texas, USA  & Clear & 9803.6469 – 9803.8163 & ruled out \\
OAUV & 0.5 & Valencia, Spain & R &  9804.3868 -- 9804.5212 &  ruled out \\
eVscope-nja & 0.1 & Renens, Vaud, Switzerland & Clear & 9804.4098 – 9804.4433 & ruled out \\
GMU$^{12}$ & 0.8 & Fairfax, VA, USA & R & 9804.6058 -- 9804.8129 & ruled out \\
Hazelwood Obs & 0.3  & Churchill, Victoria, Australia & R & 10414.1959 -- 10414.2601 & inconclusive \\
\enddata
\tablecomments{$^1$(BTJD - 2450000); $^2$Unistellar Citizen Science Network; $^3$Clear: 650 (350) nm; $^4$Las Cumbres Observatory Global Telescope \citep{Brown:2013}; $^5$Pan-STARRS $z$-short$^{3}$; $^6$Baader R 610\,nm longpass; $^7$Observatoire des Baronnies, Provencales; $^8$Frustaglia Private Observatory; $^9$The Maury Lewin Astronomical Observatory; $^{10}$Observatori Astronòmic de la Universitat de València; $^{11}$Observatori Astronomic Albanya; $^{12}$George Mason University}
\end{deluxetable*}

\subsection{NEOSSat Photometry} 

We observed a transit of \thisplanetb\ on UT 2023 March 2 with the Near Earth Object Surveillance Satellite 
\citep{Wallace2006NEOSSat, Fox2022NEOSSat, Cziranka-Crooks2023NEOSSat, Mann2024NEOSSat}. 
NEOSSat is a Canadian microsatellite orbiting the Earth in a Sun-synchronous orbit of approximately 100\,minutes. 
Though originally deployed to study near-Earth satellites, NEOSSat also performs well for follow-up observations of large exoplanets transiting bright stars. It utilizes a 15~cm f/6 telescope, no filter, a detector spectral response between 350 and 1050\,nm and a field of view of $0.86^\circ \times 0.86^\circ$.
NEOSSat observed TOI-4465 for $\sim$48~hours, interrupted by regular Earth-eclipse data gaps, resulting in $\sim$30\,minutes on target and $\sim$70\,minutes off target per orbit.
We observed from UT 2023 March 2 to 2023 March 4 (2460005.720 to 2460007.694 BJD$_{\rm TDB}$) with a 45~second cadence and 5~second exposures for a total of 1271 data points (Figure \ref{fig:transitsjointfit}). 
With these data, we achieved a 4-minute binned photometric precision of 3.8\,ppt using an aperture radius of 5 pixels ($15\arcsec$).
The light curve was detrended and normalized via a custom PCA procedure using nearby reference stars to remove instrument systematics.


\section{System Parameters from \texttt{\lowercase{juliet}} \label{sec:juliet_jointfit}}

We simultaneously modeled the TESS photometry, LCOGT, NGTS, NEOSSat and Unistellar ground-based photometry, as well as the APF, CHIRON and TRES radial velocities using the \texttt{juliet} package \citep{espinoza2019juliet}.
\texttt{juliet} employs nested sampling algorithms to efficiently search for the global posterior maximum in a given parameter space, and allows for model comparison based on Bayesian evidences. 
\texttt{juliet} combines transit modeling using \texttt{batman} \citep{kreidberg2015batman}, and RV modeling using \texttt{radvel} \citep{fulton2018radvel}. We opted to use the \texttt{dynesty} \citep{speagle2020dynesty} package to perform the posterior sampling for the global model, though a range of nested sampling algorithms are available to choose from within \texttt{juliet}.

\subsection{Transit Modeling}

\texttt{juliet} uses \texttt{batman} to model the photometric transits. 
We modeled the stellar limb darkening effect in the TESS, LCOGT, NGTS, NEOSSat and Unistellar photometry using a quadratic limb darkening model parameterized with the efficient, uninformative sampling scheme of \citet{kipping2013efficient}.
Limb darkening coefficients for each instrument were calculated using the Python Limb Darkening ToolKit \citep[PyLDTK;][]{Parviainen2015LDTK, 2015ascl.soft10003P}. PyLDTK calculates limb darkening coefficients based on PHOENIX-generated specific intensity spectra \citep{Husser2013PHOENIX} and requires inputs of stellar effective temperature, surface gravity, metallicity, and the instrument-specific bandpass.
For each photometric band -- TESS, LCOGT, NGTS, NEOSSat, and Unistellar (which is comparable to the CHEOPS band) -- we calculated quadratic coefficients ($u_1$, $u_2$) and quadratic parametrization coefficients ($q_1$, $q_2$), along with their uncertainties.
In the transit model, the TESS limb darkening coefficients ($q_{1,TESS}$, $q_{2,TESS}$) were treated as free parameters (Table \ref{tab:julietposteriors}) and assigned a normal prior (Table \ref{tab:julietpriors}), while coefficients for the other instruments were fixed to the values computed with PyLDTK (see Table \ref{tab:julietpriors} for limb darkening coefficient values).

We used a fixed dilution factor of 1 for all instruments, since there are no nearby stellar contaminating sources. Additionally, we incorporated individual instrumental offsets, and accounted for instrumental jitter by adding it in quadrature to the standard instrumental error bars.

Per the \citet{espinoza2018efficient} parameterization, we used uniform priors for the planet-to-star radius ratio, $p=R_{p}/R_{\star}$, and impact parameter, $b$, to explore the full parameter space. 

We set the stellar density as a free parameter of the transit model, and assigned it a log-uniform prior. We then obtained the scaled semi-major axis ($a/R_*$) using Kepler's third law.

\subsection{RV Modeling}

The model for the RV data included parameters for the RV semi-amplitude, $K$, orbital eccentricity, $e$, orbital argument of periastron passage, $\omega$,  individual spectrograph offsets for APF, CHIRON and TRES (systemic velocity, $\mu$), and individual instrumental jitter, $\sigma$.
We assumed uniform wide priors for the individual systemic velocities, jitter terms, RV semi-amplitude, eccentricity, and argument of periastron passage.
We measured a radial velocity semi-amplitude of $K = 277.19^{+2.37}_{-1.90}~\ms$ for \thisplanetb.

We modeled the TOI-4465 RVs including and excluding the CORALIE RV data. The CORALIE data were ultimately excluded from the final RV model because of their large errorbars, which increased parameter uncertainties and added noise without enhancing model precision. The limited number of CORALIE data points, along with an incomplete planetary orbit, likely contributed to this effect by introducing additional free parameters that reduced the model’s accuracy. \\

We show the final RV model of the joint fit in Figure \ref{fig:RVsjointfit}, and transit models of the joint fit in Figures \ref{fig:tessdataphased}, \ref{fig:transitsjointfit} and \ref{fig:unistellartransitsjointfit}, based on the posterior sampling. The posterior parameters of our joint fit are listed in Table \ref{tab:julietposteriors}, and the obtained posterior probabilities are shown in Figure \ref{fig:cornerposteriors}. The derived planetary parameters of \thisplanetb\ based on the posteriors of the joint fit are listed in Table \ref{tab:julietderived}. Priors for our joint fit are listed in Table \ref{tab:julietpriors}. The Unistellar photometry priors and posteriors are listed in the Appendix in Table \ref{tab:julietpriors} and Table \ref{tab:unistellarposteriors}, respectively.

\startlongtable
\begin{deluxetable*}{llccc}
\tabletypesize{\small}
\tablewidth{\linewidth}
\tablecaption{Median values and 68\% confidence interval for posterior parameters from joint photometric and radial velocity \texttt{juliet} analysis for the TOI-4465 system.}
\tablehead{\colhead{Parameter} & \colhead{Units} & \colhead{Values}}
\startdata
\smallskip\\\multicolumn{2}{l}{Stellar Parameters:}\\
$\rho_*$ & Stellar density (g cm$^{-3}$) & $1.30^{+0.41}_{-0.44}$\\
\smallskip\\\multicolumn{2}{l}{Planet Parameters:}&TOI-4465 b\smallskip\\
$P$ &Period (days) &$101.94054^{+0.00040}_{-0.00036}$\\
$T_0$ & Time of transit center (\bjdtdb) &$2459395.1256\pm0.0006$\\
$r_1$ & Parametrization of \citet{espinoza2018efficient} for $b$ & $0.39^{+0.05}_{-0.03}$\\
$r_2$ & Parametrization of \citet{espinoza2018efficient} for $R_p/R_*$ & $0.1268\pm0.0004$\\
$K$ & RV semi-amplitude (\ms) & $277.19^{+2.37}_{-1.90}$\\
$e$ & Eccentricity & $0.24\pm0.01$\\
$\omega$ & Argument of periastron passage ($^\circ$) & $279.74^{+0.58}_{-0.57}$\\
\smallskip\\\multicolumn{2}{l}{Photometry Parameters:}&TESS\\
$M$ & Relative flux offset & $0.0000$ \\
$\sigma$ &  Jitter term for light curve (ppm) & $554\pm11$\\
$q_{1}$ & Quadratic limb darkening parametrization \citep{kipping2013efficient} & $0.31\pm0.02$ \\
$q_{2}$ & Quadratic limb darkening parametrization \citep{kipping2013efficient} & $ 0.39\pm0.01$ \smallskip\\
\multicolumn{2}{l}{}&LCOGT&NGTS&NEOSSat\\
$M$ & Relative flux offset & $-0.0008$ & $-0.0022$ & $-0.0015$\\
$\sigma$ &  Jitter term for light curve (ppm) & $3557^{+129}_{-125}$ & $10^{+88}_{-9}$ & $2170^{+421}_{-489}$\\
\smallskip\\\multicolumn{2}{l}{RV Parameters:}\smallskip\\
$\mu_{APF}$ & Systemic velocity for APF (\ms) & $45.23^{+1.72}_{-1.52}$ \\
$\sigma_{APF}$ &  Jitter term for APF (\ms) & $9.90^{+1.37}_{-1.35}$ \\
$\mu_{CHIRON}$ & Systemic velocity for CHIRON (\ms) & $-11782.8^{+3.1}_{-3.2}$ \\
$\sigma_{CHIRON}$ &  Jitter term for CHIRON (\ms) & $9.1^{+3.4}_{-2.9}$ \\
$\mu_{TRES}$ & Systemic velocity for TRES (\ms) & $298.91^{+10.24}_{-9.99}$ \\
$\sigma_{TRES}$ &  Jitter term for TRES (\ms) & $29.70^{+12.04}_{-9.91}$ \\
\label{tab:julietposteriors} 
\enddata
\end{deluxetable*}

\clearpage
\startlongtable
\begin{deluxetable*}{llccc}
\tabletypesize{\small}
\tablewidth{\linewidth}
\tablecaption{Median values and 68\% confidence interval for derived parameters from joint photometric and radial velocity \texttt{juliet} analysis for TOI-4465 b.}
\tablehead{\colhead{Parameter} & \colhead{Units} & \colhead{Values}}
\startdata
\smallskip\\\multicolumn{2}{l}{Derived Transit Parameters:}&\smallskip\\
$p=R_p/R_*$ & Radius of planet in stellar radii & $0.1268\pm0.0004$\\
$a/R_*$ & Semi-major axis in stellar radii  & $88.4\pm4.3$\\
$b=(a/R_*)cos(i_p)$ & Transit impact parameter & $0.080^{+0.083}_{-0.051}$\\
$i_p$ & Inclination (degrees) & $89.95^{+2.71}_{-4.44}$\\
\smallskip\\\multicolumn{2}{l}{}&TESS\\
$u_{1}$ & Linear limb darkening coefficient & $0.43\pm0.02$ \\
$u_{2}$ & Quadratic limb darkening coefficient & $0.12\pm0.02$ \\
\smallskip\\\multicolumn{2}{l}{Derived Planetary Parameters:}&TOI-4465 b\smallskip\\
$R_p$ & Radius (\rj) & $1.25^{+0.08}_{-0.07}$\\
$M_p$ & Mass (\mj) & $5.89 \pm 0.26$\\
$\rho_p$ & Density (g cm$^{-3}$) &$3.73 \pm 0.53$\\
$a$ & Semi-major axis (AU) &$0.416\pm0.014$\\
$T_{eq:peri.}$ & Equilibrium temperature at periastron (K) & $478 \pm 15$\\
$T_{eq:apa.}$ & Equilibrium temperature at apastron (K) & $375 \pm 13$\\
$g_p$ & Surface gravity (m~s$^{-2}$) & $93.2\pm11.9$\\
\label{tab:julietderived} 
\enddata
\end{deluxetable*}


\section{Discussion \label{sec:discussion}}

The \thisstar\ system consists of a G dwarf host star with a long-period giant planet, \thisplanetb\ (see Table \ref{tab:julietderived}) which has a mass of  $M_{p}$ = \plmassunc\ and radius of $R_{p}$ = \plradunc\, on a mildly eccentric orbit ($e=0.24\pm0.01$) with a period of $\sim$102~days. We derived a bulk density of $\rho$ = \plrhounc\ and an equilibrium temperature range, assuming a zero albedo, of $T_{eq} = 375-478$~K.

With the physical and orbital parameters of TOI-4465~b well constrained, we place the planet in the context of other giant planet discoveries, and discuss the potential for planetary atmospheric characterization, the planet's bulk heavy element content, and prospects for obliquity measurement.

\subsection{TOI-4465~b in the Population Context}

\begin{figure*}
    \centering
    \includegraphics[width=\linewidth,keepaspectratio]{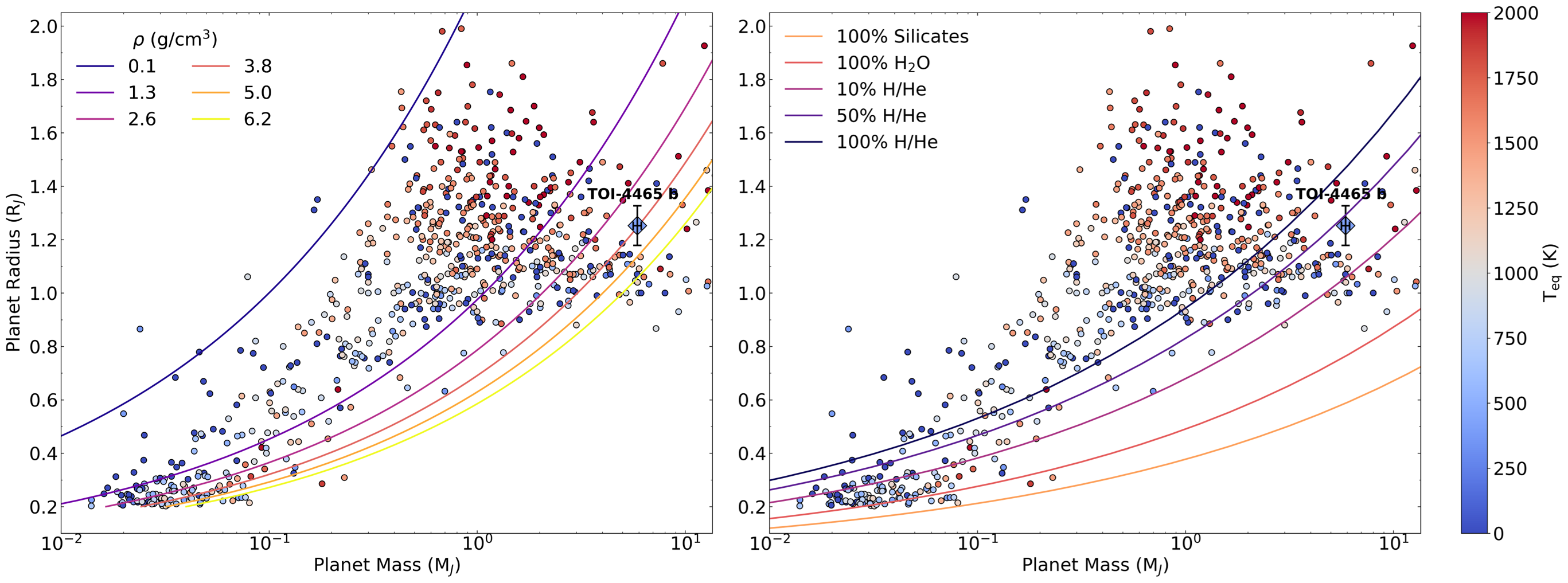}
\caption{Mass-radius diagram for confirmed giant planets ($R_p > 0.2~\rj$, $M_p < 13~\mj$) with measured mass and radius uncertainties below 20\%. Planets are colored according to their average equilibrium temperature (calculated assuming zero albedo and efficient heat redistribution). \emph{Left}: mass–radius diagram with curves of constant bulk densities ranging from 0.1~g cm$^{-3}$ (dark blue) to 6.2~g cm$^{-3}$ (yellow). \emph{Right}: mass–radius diagram with composition curves from theoretical models in \citet{fortney2007planetary}. From bottom curve to top: 100\% silicates (light orange), 100\% H$_2$O (dark orange), 10\% H/He atmosphere (magenta), 50\% H/He (purple), and 100\% H/He (dark blue). TOI-4465~b has a bulk density of \plrho\, and is one of very few known giant planets that are large, massive, dense, and temperate. \label{fig:massradius}}
\end{figure*}

\begin{figure}
    \centering
    \includegraphics[width=\linewidth,keepaspectratio]{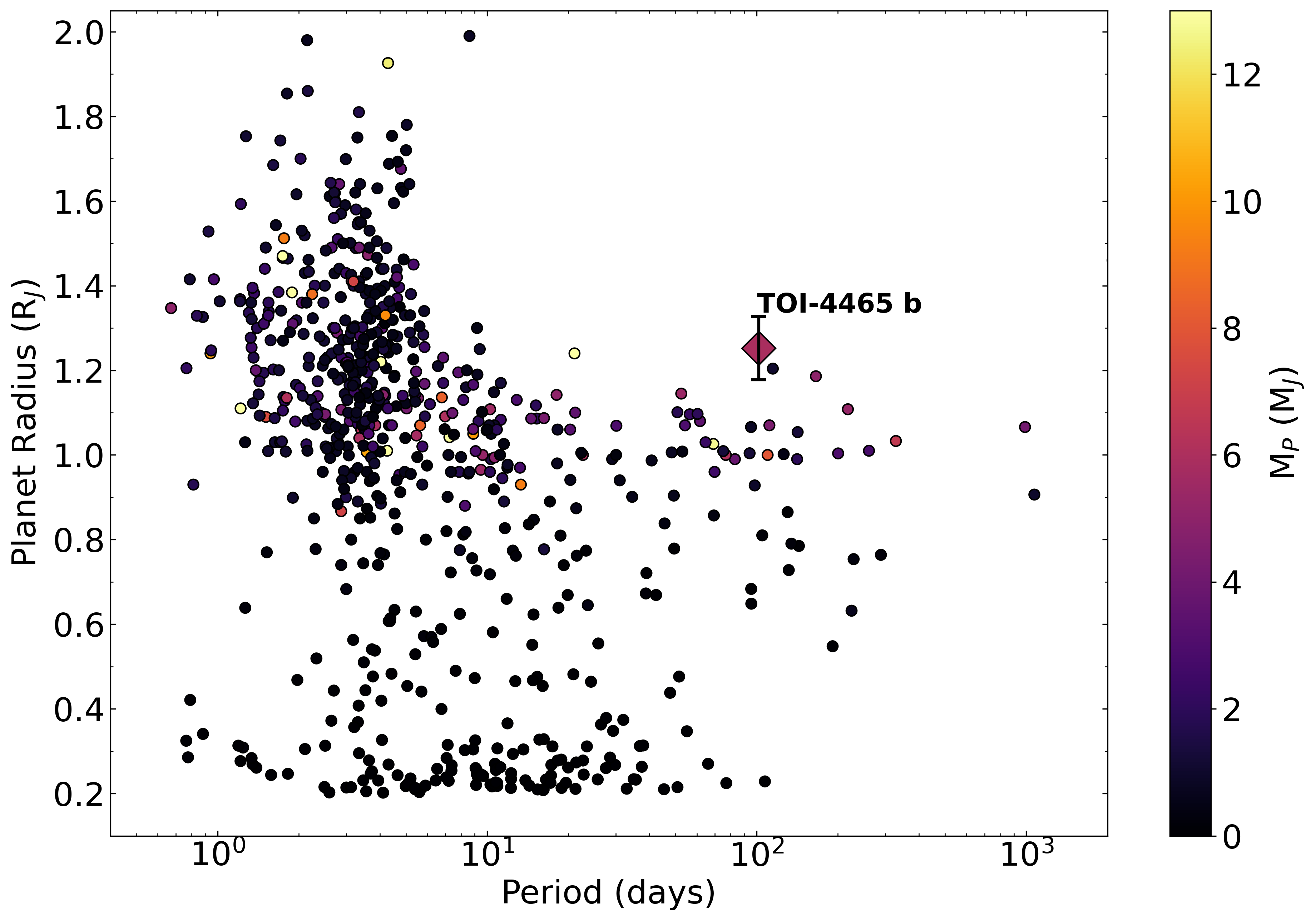}
\caption{Period-radius plot for confirmed giant planets ($R_p > 0.2~\rj$, $M_p < 13~\mj$) with measured mass and radius uncertainties below 20\%. Planets are colored according to their mass. TOI-4465~b has the largest radius among giant planets with $P>100$~days, and the second most massive planet with $R_p > 1.2~\rj$ and $P>10$~days. \label{fig:periodradius}}
\end{figure}

\begin{figure}
    \centering
    \includegraphics[width=\linewidth,keepaspectratio]{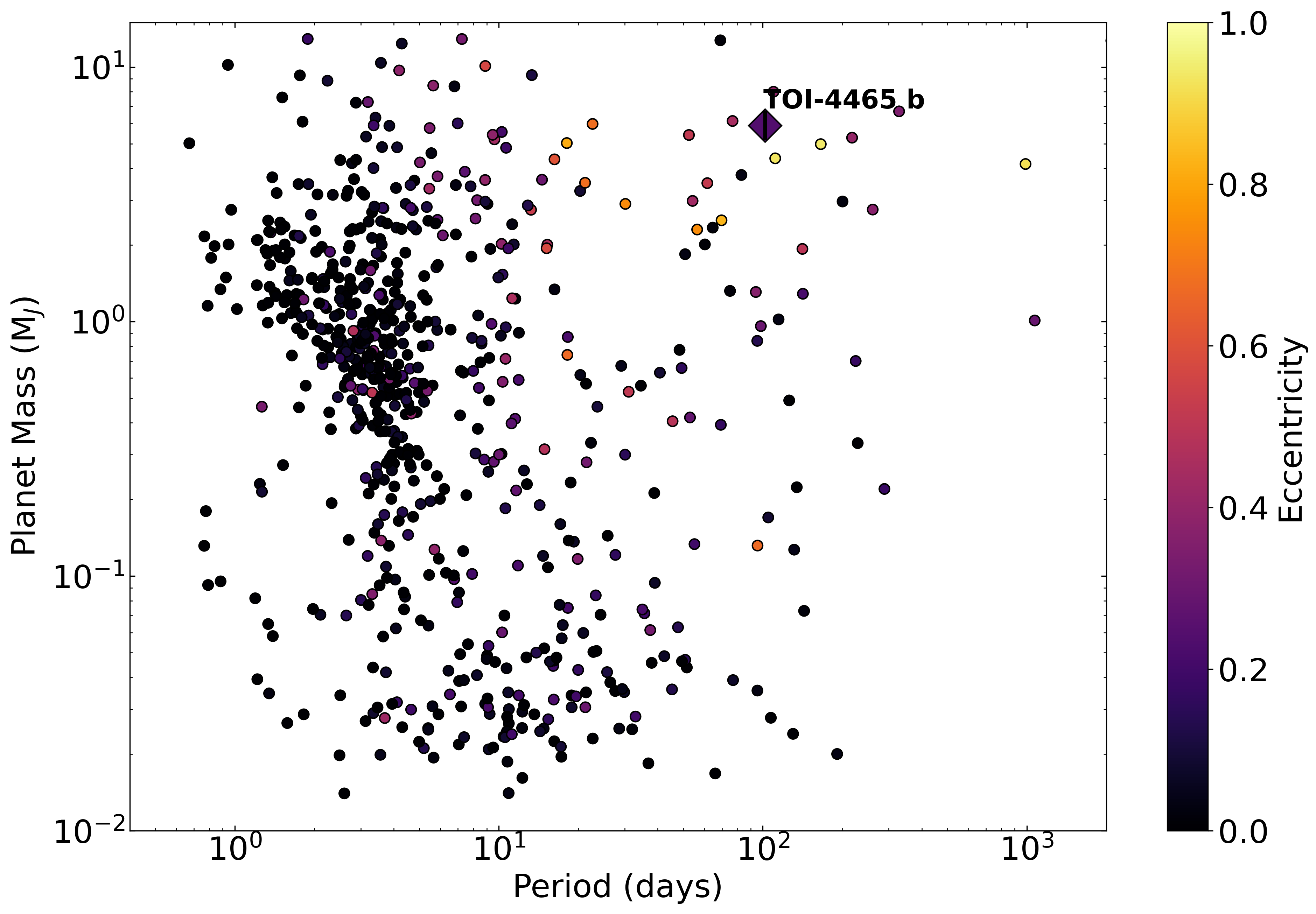}
\caption{Period-mass plot for confirmed giant planets ($R_p > 0.2~\rj$, $M_p < 13~\mj$) with measured mass and radius uncertainties below 20\%. Planets are colored according to their orbital eccentricity. Planets at longer orbital periods generally trend towards higher eccentricities. TOI-4465~b's eccentricity is comparatively mild for its mass and period.  \label{fig:periodmass}}
\end{figure}

We place \thisplanetb\ in the context of other giant planet discoveries by constructing a sample of confirmed transiting giant exoplanets ($R_p > 0.2~\rj$, $M_p < 13~\mj$) with precise mass and radius measurements (uncertainties $<$ 20\%, or precision $> 5\sigma$) from the NASA Exoplanet Archive\footnote{\href{https://exoplanetarchive.ipac.caltech.edu/}{https://exoplanetarchive.ipac.caltech.edu/}; Accessed May 1, 2025}. We present this giant planet sample in three parameter spaces: mass-radius (Figure \ref{fig:massradius}), period-radius (Figure \ref{fig:periodradius}), and period-mass (Figure \ref{fig:periodmass}).

TOI-4465~b is in an underpopulated region of mass-radius space (Figure \ref{fig:massradius}), and is one of very few known giant planets that are large, massive, dense, and temperate. 

In period-radius space (Figure \ref{fig:periodradius}), TOI-4465~b has the largest radius among giant planets with $P>100$~days. It is also the second most massive planet with $R_p > 1.2~\rj$ and $P>10$~days. TOI-4465~b's long orbital period results in a low insolation ($\sim 5 S_\earth$), preventing significant radius inflation -- the phenomenon where a planet's radius becomes much larger due to intense heating. Radius inflation can complicate efforts to determine a planet's bulk composition, as the radius inflation factor may not be known, which adds uncertainty to interpretations of the planet's internal structure. In contrast, the lack of significant radius inflation in the case of TOI-4465~b allows for more straightforward determination of its bulk properties, providing a clearer baseline for studies of its formation processes and atmospheric structure.

In period-mass space (Figure \ref{fig:periodmass}), TOI-4465~b is among the most massive, long-period giant planets, and appears to have a relatively low eccentricity compared to its immediate neighbors, though we note that transit detection bias likely favors the discovery of eccentric long-period planets due to their higher transit probability at periastron.
Systems with giant planets can often exhibit a wide variety of orbital eccentricities, often the result of planet-planet and/or planet-disk interactions that result in angular momentum exchange \citep{raymond2010,kane2014,clement2021}. In some cases, the gravitational interactions can escalate to the point of planet-planet scattering that ejects one or more planets from the system, where the angular momentum of the lost planet(s) is preserved within the high eccentricity of the remaining planets \citep{carrera2019,kane2023}. For these reasons, planets at longer orbital periods generally trend towards higher eccentricities \citep{kane2012,wittenmyer2020,kane2024}, potentially influencing the volatile inventory of inner planets \citep{raymond2017,marov2018,kane2024}, and emphasizing TOI-4465~b's distinctive dynamical properties. Indeed, \citet{kane2024} found that the median eccentricity of giant planets interior to the snow line (excluding circularized hot Jupiters) is $\sim$0.24. Therefore, the eccentricity of TOI-4465~b is consistent with the known population of temperate giant planets. Although the data analysis described in Section~\ref{sec:juliet_jointfit} prefers a single planet solution, the present eccentricity may be indicative of current angular momentum exchange with a yet unseen outer planet, or the remnants of past disk interaction during migration.

\subsection{Prospects for Atmospheric Characterization}

We calculated the transmission spectroscopy metric (TSM) and emission spectroscopy metric (ESM), defined in \citet{kempton2018framework}, to determine TOI-4465~b's potential for atmospheric characterization We found a TSM range (between apoastron and periastron) of TSM~$=9.5(\pm1.9) - 12.2(\pm2.4)$, and an ESM range of ESM~$=2.9(\pm0.4) - 9.0(\pm1.2)$ for TOI-4465~b.

TOI-4465~b joins a small population of long-period, cool/temperate transiting giant planets. There are 21 confirmed planets\footnote{\dataset[NASA Exoplanet Archive]{https://doi.org/10.26133/NEA12}; Accessed May 1, 2025} with $P >100$~days, and mass and radius precision $>5\sigma$. The range of radii, masses, and average equilibrium temperatures for this population of long-period planets are $0.23 \leq R_p \leq 1.20 \rj$, $0.02 \leq M_p \leq 8.0 \mj$, and $134 \leq T_{eq} < 500$~K respectively. 

We calculated a TSM range for the population of TSM~$= 0.5 - 128$. Due to its high mass, and hence high surface gravity, TOI-4465~b is likely not a good candidate for atmospheric characterization via transmission spectroscopy. However, TOI-4465~b has the third highest ESM of all the planets in the population described above. Aside from HD~80606~b\footnote{HD 80606 b has a ESM range of ESM~$=15-231$ due to its extreme eccentricity ($e=0.93)$ \citep{rosenthal2021california}.} and TIC~241249530~b\footnote{TIC~241249530~b has a ESM range of ESM~$=0.25-57$ due to its extreme eccentricity ($e=0.94)$ \citep{gupta2024hot}.}, all the planets in the population have an average ESM $\leq3$. Thus, TOI-4465~b is one of the best long-period, well-constrained, temperate planets available for emission spectroscopy studies.

The atmospheres of temperate giant planets provide a window into the chemistry of cooler environments that can bridge the gap between hot Jupiters and cold solar system gas giants. Within TOI-4465~b's temperature range ($T_{eq} = 375-478$~K), disequilibrium chemistry via shifts in CO/CH$_4$ and N$_2$/NH$_3$ ratios driven by vertical transport could be observed in both transmission and emission spectroscopy \citep{fortney2020beyond}. Additionally, transitions in nitrogen chemistry and sulfur hazes are also expected to occur in TOI-4465~b's temperature range \citep{ohno2023nitrogen, gao2017sulfur}.

\subsection{Bulk Heavy-element Content}

Following \citet{thorngren2016mass}, we calculated the heavy-element enrichment of TOI-4465~b relative to its host star. We found a bulk metallicity (heavy-element mass fraction) of $Z_p = 0.091 \pm 0.017$ for TOI-4465~b\footnote{$Z_p$ is an upper limit on the atmospheric metal fraction because for planets more massive than Saturn, the heavy elements are expected to be in the envelope rather than the core.}, which corresponds to a planet heavy-element mass of $M_z = 171 \pm 32~\Mearth$.
We approximate the stellar heavy-element mass, $Z_*$, using $Z_* = 0.0142 \times 10^{\mathrm{[Fe/H]}}$, assuming stellar metallicity scales with iron abundance, resulting in $Z_* = 0.0122 \pm 0.0004$. 
The heavy-element enrichment of TOI-4465~b is then $Z_p/Z_* = 7.5 \pm 1.4$. TOI-4465~b is consistent with the mass-metallicity correlation measured by \citet{thorngren2016mass} for planets with $P < 100$~days.

Studies have found a higher occurrence rate for close-in giant planets orbiting higher metallicity host stars \citep[e.g.][]{petigura2018california, osborn2020investigating}. Some recent studies point to a potential separation in the populations of giant planets at around 4\mj, where more massive giant planets orbit lower metallicity host stars \citep[e.g.][]{santos2017observational, maldonado2019connecting, goda2019multiple}. This potential separation in giant planet mass with host star metallicity could indicate different planet formation processes for the different populations: lower mass giant planets form via core accretion \citep{pollack1996formation}, and higher mass giant planets form via disk instability \citep{boss1997giant}. However, \citet{adibekyan2019heavy} found no clear demarcation between these formation pathways with giant planet mass, and instead proposed that high mass planets might form through various processes depending on their specific initial conditions. 

TOI-4465 b is the most massive transiting giant planet ($P > 10$ days) orbiting a sub-solar metallicity host star ($\mathrm{[Fe/H]} = -0.060 \pm 0.016$~dex). While TOI-4465~b is significantly metal-enriched relative to its host star, this enrichment aligns with predictions from core accretion theory with late-stage accretion of icy planetesimals, which can lead to enrichments up to, and sometimes greater than, 10 \citep[e.g.][]{mordasini2016imprint, thorngren2016mass}.  Late-stage accretion of planetesimals from the protoplanetary disk can also lead to metal-enriched planets formed via disk instability \citep[e.g.][]{helled2006planetesimal}. Consequently, TOI-4465~b does not provide conclusive evidence for different formation processes between high-mass and low-mass giant planets, but it does represent a valuable observational data point as the sample of long-period giant planets continues to grow.

\subsection{RM Effect \& Obliquity}

By measuring the Rossiter-McLaughlin (RM) effect \citep{1924ApJ....60...22M, 1924ApJ....60...15R} for the system, we can determine the sky-projected obliquity, providing valuable insights into the formation and evolution of TOI-4465~b. A misaligned orbit could suggest the presence of an additional planet in the system responsible for perturbing the orbit of TOI-4465~b. Conversely, an aligned orbit could indicate that the orbit of TOI-4465~b is a result of disk migration \citep{queloz2000detection, winn2010hot, triaud2010spin}.

Using Equation 1 from \citet{2018haex.bookE...2T}, and the values for $R_p/R_*$ and $b$ (Table \ref{tab:julietderived}), and $v$sin$i$ (upper limit; Section \ref{subsubsec:coralie}), we estimate an RM effect semi-amplitude upper limit for TOI-4465~b of 42~\ms, which is well within the detection capabilities of current facilities, and would make TOI-4465~b one of the longest-period planets with a measured obliquity. However, we note that the long period and transit duration of TOI-4465~b would make this measurement challenging, likely requiring multiple transit observations.


\section{Conclusions\label{sec:conclusions}}

We report the discovery and confirmation of TOI-4465~b, a transiting, long-period, temperate giant planet orbiting a G dwarf star. The planet was initially identified as a single-transit event in data from TESS Sector 40. Follow-up RV observations of the star with APF, CHIRON and TRES determined the orbital period ($\sim$102~days), mass ($M_{p} = $ \plmassunc) and eccentricity ($e=0.24\pm0.01$) of TOI-4465~b. The planet’s equilibrium temperature ranges from 375 – 478~K, given its mild orbital eccentricity. A global ground-based photometry campaign was initiated to observe another transit of TOI-4465~b after the RV period determination. Notably, this campaign included contributions from 24 citizen scientists. Photometric observations from TESS, LCOGT, NGTS, NEOSSat and the Unistellar Citizen Science Network were used to measure the planet's radius ($R_{p} = $ \plradunc). TOI-4465~b is one of the best long-period ($P >$ 100 days) giant planets available for atmospheric emission spectroscopy studies. We estimate the bulk heavy-element content of TOI-4465~b and find that the planet is metal-enriched relative to its host star by a factor of $7.5 \pm 1.4$. Additionally, we find that TOI-4465~b is a good candidate for obliquity measurement via the RM effect. Future observations of the spin-orbit angle and the atmosphere of the planet can provide insights into the formation, evolution and composition of the system. Confirming single-transit events is the leading strategy for discovering long-period exoplanets, which is crucial for better understanding the currently sparsely populated outer regions of planetary systems.

\section*{Acknowledgments}

Z.E. acknowledges support from the TESS Guest Investigator Program grant 80NSSC23K0769. 
D.D. acknowledges support from the TESS Guest Investigator Program grant 80NSSC23K0769, and from the NASA Exoplanet Research Program grant 80NSSC20K0272.
D.R.C. acknowledges partial support from NASA Grant 18-2XRP18 2-0007. 
We thank Ken and Gloria Levy, who supported the construction of the Levy Spectrometer on the Automated Planet Finder. We thank the University of California and Google for supporting Lick Observatory and the UCO staff for their dedicated work scheduling and operating the telescopes of Lick Observatory.
P.A.D. acknowledges support by a 51 Pegasi b Postdoctoral Fellowship from the Heising-Simons Foundation and by a National Science Foundation (NSF) Astronomy and Astrophysics Postdoctoral Fellowship under award AST-1903811.
L.A.S. and T.M.E. were partially supported during this work by the NASA Citizen Science Seed Funding Program via grant number 80NSSC22K1130 and the NASA Exoplanets Research Program via grant 80NSSC24K0165. Those NASA grants also support the UNITE (Unistellar Network Investigating TESS Exoplanets) program, under the auspices of which the Unistellar data were collected.
The Unistellar Network also acknowledges financial support during its foundational phase from Frédéric Gastaldo and the Gordon and Betty Moore Foundation.
We thank Niniane Leroux for her contributions to observations as a Unistellar Citizen Scientist.
This work makes use of observations from the LCOGT network. Part of the LCOGT telescope time was granted by NOIRLab through the Mid-Scale Innovations Program (MSIP). MSIP is funded by NSF.
This research has made use of the Exoplanet Follow-up Observation Program (ExoFOP; DOI: 10.26134/ExoFOP5) website, which is operated by the California Institute of Technology, under contract with the National Aeronautics and Space Administration under the Exoplanet Exploration Program.
This paper made use of data collected by the TESS mission and are publicly available from the Mikulski Archive for Space Telescopes (MAST) operated by the Space Telescope Science Institute (STScI). Funding for the TESS mission is provided by NASA's Science Mission Directorate. We acknowledge the use of public TESS data from pipelines at the TESS Science Office and at the TESS Science Processing Operations Center. Resources supporting this work were provided by the NASA High-End Computing (HEC) Program through the NASA Advanced Supercomputing (NAS) Division at Ames Research Center for the production of the SPOC data products. KAC and CNW acknowledge support from the TESS mission via subaward s3449 from MIT. 
Some of the observations in this paper made use of the High-Resolution Imaging instrument ‘Alopeke and were obtained under Gemini LLP Proposal Number: GN/S-2021A-LP-105. ‘Alopeke was funded by the NASA Exoplanet Exploration Program and built at the NASA Ames Research Center by Steve B. Howell, Nic Scott, Elliott P. Horch, and Emmett Quigley. Alopeke was mounted on the Gemini North telescope of the international Gemini Observatory, a program of NSF’s OIR Lab, which is managed by the Association of Universities for Research in Astronomy (AURA) under a cooperative agreement with the National Science Foundation. on behalf of the Gemini partnership: the National Science Foundation (United States), National Research Council (Canada), Agencia Nacional de Investigación y Desarrollo (Chile), Ministerio de Ciencia, Tecnología e Innovación (Argentina), Ministério da Ciência, Tecnologia, Inovações e Comunicações (Brazil), and Korea Astronomy and Space Science Institute (Republic of Korea).
Some observations in this work were obtained at the Southern Astrophysical Research (SOAR) telescope, which is a joint project of the Minist\'{e}rio da Ci\^{e}ncia, Tecnologia e Inova\c{c}\~{o}es (MCTI/LNA) do Brasil, the US National Science Foundation’s NOIRLab, the University of North Carolina at Chapel Hill (UNC), and Michigan State University (MSU)."
I.A.S. (SAI speckle imaging) acknowledges the support of M.V. Lomonosov Moscow State University Program of Development.
Part of this research was carried out at the Jet Propulsion Laboratory, California Institute of Technology, under a contract with the National Aeronautics and Space Administration (NASA). 
This work has been carried out within the framework of the NCCR PlanetS supported by the Swiss National Science Foundation under grants 51NF40{\_}182901 and 51NF40{\_}205606.
This work is based in part on data collected under the NGTS project at the ESO La Silla Paranal Observatory. The NGTS facility is operated by a consortium institutes with support from the UK Science and Technology Facilities Council (STFC) under projects ST/M001962/1, ST/S002642/1 and ST/W003163/1.
The contributions at the Mullard Space Science Laboratory by E.M.B. have been supported by STFC through the consolidated grant ST/W001136/1.
The postdoctoral fellowship of K.B. is funded by F.R.S.-FNRS grant T.0109.20 and by the Francqui Foundation.

\facilities{TESS, LCOGT, NGTS, Unistellar, NEOSSat, APF, CHIRON, TRES, CORALIE, Keck I (HIRES), Gemini-North (‘Alopeke), SOAR, SAI, Palomar Hale (PHARO)}

\software{AstroImageJ \citep{Collins:2017}, TAPIR \citep{Jensen:2013}, juliet \citep{espinoza2019juliet}, dynesty \citep{speagle2020dynesty}. Lightkurve \citep{2018lightkurve}, PyLDTK \citep{2015ascl.soft10003P}}

\bibliographystyle{aasjournal}
\bibliography{references}

\clearpage
\appendix
\renewcommand{\thefigure}{A\arabic{figure}}
\renewcommand{\thetable}{A\arabic{table}}
\setcounter{figure}{0}
\setcounter{table}{0}

\begin{figure*}[htb!]
    \centering
    \includegraphics[width=\textwidth,keepaspectratio]{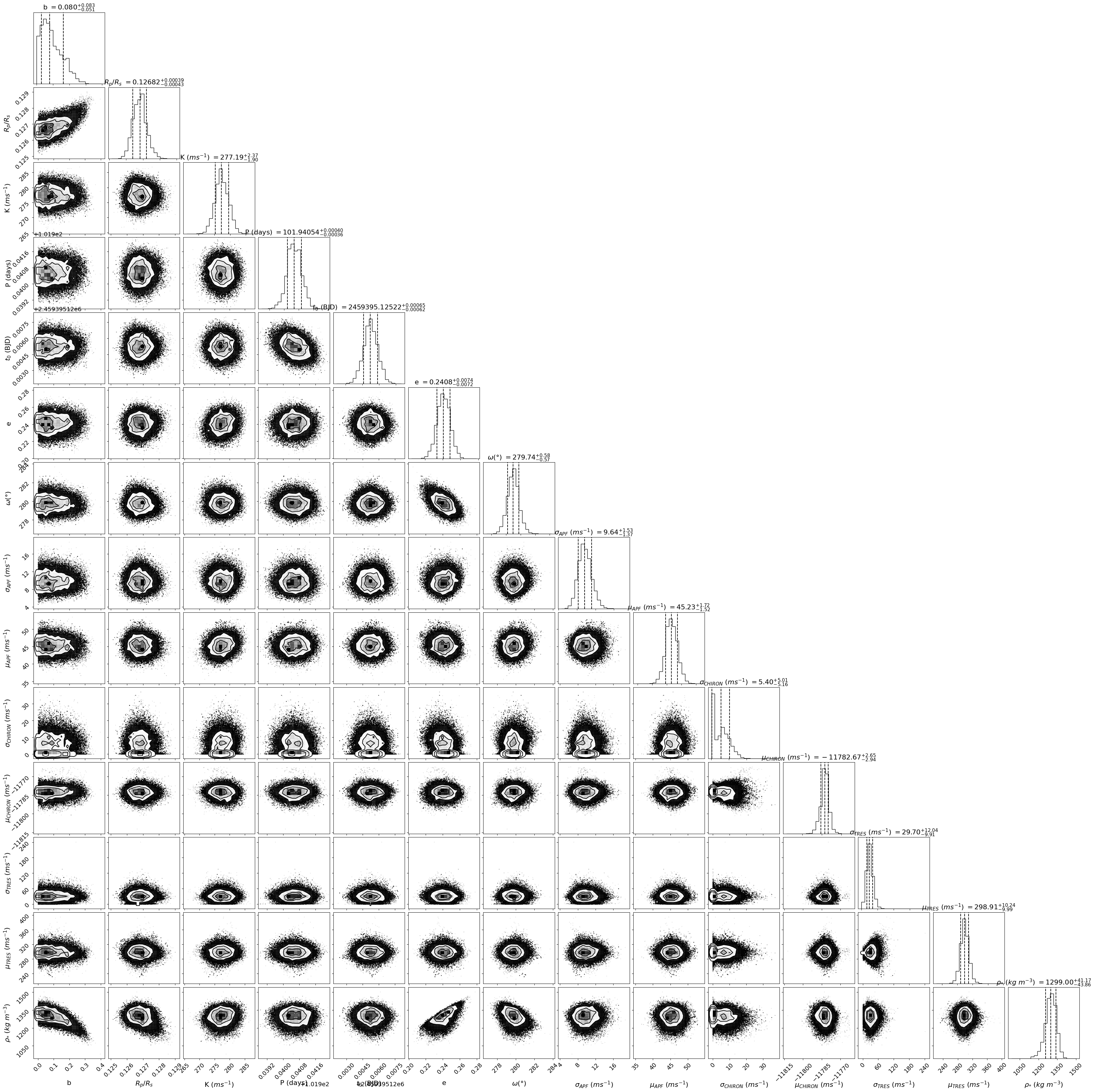}
\caption{Posterior distribution for the joint (photometric + radial velocity) model parameters derived with \texttt{juliet}.  \label{fig:cornerposteriors}}
\end{figure*}

Here we display the posterior distributions and priors for our joint analysis of the TOI-4465 system. Posterior distributions for the \texttt{juliet} joint model parameters are shown in Figure \ref{fig:cornerposteriors}. Priors used in the joint analysis are listed in Table \ref{tab:julietpriors}. The Unistellar photometry posteriors are listed in Table \ref{tab:unistellarposteriors}.

\startlongtable
\begin{deluxetable*}{llccc}
\tabletypesize{\small}
\tablewidth{\linewidth}
\tablecaption{Priors used in our joint analysis of the TOI-4465 system with \texttt{juliet}. The prior labels of $\mathcal{N}$, $\mathcal{U}$, and $\mathcal{LU}$ represent normal, uniform, and log-uniform distributions, respectively, where $\mathcal{N}(\mu, \sigma^2)$ is a normal distribution of the mean $\mu$ and variance $\sigma^2$, and $\mathcal{U}(a, b)$ and
$\mathcal{LU}(a, b)$ are uniform and log-uniform distributions between a and b.}
\tablehead{\colhead{Parameter} & \colhead{Prior} & \colhead{Units and Description}}
\startdata
\\\multicolumn{2}{l}{Stellar Parameters:}&\smallskip\\
$\rho_*$ & ~~~~~~~~~~~~~~~~~~~~~~~~$\mathcal{LU}$($10^2, 10^5)$ & Stellar density of TOI-4465 (kg m$^{-3})$\\
\\\multicolumn{2}{l}{Planet Parameters:}& TOI-4465 b\smallskip\\
$P_{b}$ & ~~~~~~~~~~~~~~~~~~~~~~~~$\mathcal{N}$($101.0, 1.0)$ & Planet period (days)\\
$T_{0,b}$ & ~~~~~~~~~~~~~~~~~~~~~~~~$\mathcal{N}$($2459395.1250, 1.0)$ & Time of transit center (days)\\
$r_{1,b}$ & ~~~~~~~~~~~~~~~~~~~~~~~~$\mathcal{U}$($0, 1)$ & Parametrization of \citet{espinoza2018efficient} for $p$ and $b$\\
$r_{2,b}$ & ~~~~~~~~~~~~~~~~~~~~~~~~$\mathcal{U}$($0, 1)$ & Parametrization of \citet{espinoza2018efficient} for$p$ and $b$\\
$K_b$ & ~~~~~~~~~~~~~~~~~~~~~~~~$\mathcal{U}$($0, 20000.0)$ & RV semi-amplitude (\ms)\\
$e_b$ & ~~~~~~~~~~~~~~~~~~~~~~~~$\mathcal{U}$($0, 1)$ & Orbital eccentricity\\
$\omega_b$ & ~~~~~~~~~~~~~~~~~~~~~~~~$\mathcal{U}$($0, 360)$ & Periastron angle (deg)\\
\\\multicolumn{2}{l}{Photometry Parameters (TESS, LCOGT, NGTS, NEOSSat):}\\
$D_{TESS}$ & ~~~~~~~~~~~~~~~~~~~~~~~~1.0 (fixed) & Dilution factor for \TESS\\
$M_{TESS}$ & ~~~~~~~~~~~~~~~~~~~~~~~~$\mathcal{N}$($0, 0.1)$ & Relative flux offset for \TESS\\
$\sigma_{TESS}$ & ~~~~~~~~~~~~~~~~~~~~~~~~$\mathcal{LU}$($0, 1000)$ & Jitter term for \TESS\ light curve (ppm)\\
$q_{1,TESS}$ & ~~~~~~~~~~~~~~~~~~~~~~~~$\mathcal{N}$($0.3, 0.1)$ & Quadratic limb darkening parametrization \citep{kipping2013efficient} \\
$q_{2,TESS}$ & ~~~~~~~~~~~~~~~~~~~~~~~~$\mathcal{N}$($0.4, 0.1)$ & Quadratic limb darkening parametrization \citep{kipping2013efficient}\\
$D_{LCOGT}$ & ~~~~~~~~~~~~~~~~~~~~~~~~1.0 (fixed) & Dilution factor for LCOGT\\
$M_{LCOGT}$ & ~~~~~~~~~~~~~~~~~~~~~~~~$\mathcal{N}$($0, 0.1)$ & Relative flux offset for LCOGT\\
$\sigma_{LCOGT}$ & ~~~~~~~~~~~~~~~~~~~~~~~~$\mathcal{LU}$($0.1, 10^5)$ & Jitter term for LCOGT light curve (ppm)\\
$q_{1,LCOGT}$ & ~~~~~~~~~~~~~~~~~~~~~~~~0.2497 (fixed) & Quadratic limb darkening parametrization \citep{kipping2013efficient} \\
$q_{2,LCOGT}$ & ~~~~~~~~~~~~~~~~~~~~~~~~0.3775 (fixed) & Quadratic limb darkening parametrization \citep{kipping2013efficient}\\
$D_{NGTS}$ & ~~~~~~~~~~~~~~~~~~~~~~~~1.0 (fixed) & Dilution factor for NGTS\\
$M_{NGTS}$ & ~~~~~~~~~~~~~~~~~~~~~~~~$\mathcal{N}$($0, 0.1)$ & Relative flux offset for NGTS\\
$\sigma_{NGTS}$ & ~~~~~~~~~~~~~~~~~~~~~~~~$\mathcal{LU}$($0.1, 10^5)$ & Jitter term for NGTS light curve (ppm)\\
$q_{1,NGTS}$ & ~~~~~~~~~~~~~~~~~~~~~~~~0.3732 (fixed) & Quadratic limb darkening parametrization \citep{kipping2013efficient} \\
$q_{2,NGTS}$ & ~~~~~~~~~~~~~~~~~~~~~~~~0.3976 (fixed) & Quadratic limb darkening parametrization \citep{kipping2013efficient} \\
$D_{NEOSSat}$ & ~~~~~~~~~~~~~~~~~~~~~~~~1.0 (fixed) & Dilution factor for NEOSSat\\
$M_{NEOSSat}$ & ~~~~~~~~~~~~~~~~~~~~~~~~$\mathcal{N}$($0, 0.1)$ & Relative flux offset for NEOSSat\\
$\sigma_{NEOSSat}$ & ~~~~~~~~~~~~~~~~~~~~~~~~$\mathcal{LU}$($0.1, 10^5)$ & Jitter term for NEOSSat light curve (ppm)\\
$q_{1,NEOSSat}$ & ~~~~~~~~~~~~~~~~~~~~~~~~0.4252 (fixed) & Quadratic limb darkening parametrization \citep{kipping2013efficient} \\
$q_{2,NEOSSat}$ & ~~~~~~~~~~~~~~~~~~~~~~~~0.4097 (fixed) & Quadratic limb darkening parametrization \citep{kipping2013efficient} \\
\vspace{-0.25cm}
\\\multicolumn{2}{l}{Unistellar Photometry Parameters (for all 13 instruments):}&\smallskip\\
$D_{Unistellar}$ & ~~~~~~~~~~~~~~~~~~~~~~~~1.0 (fixed) & Dilution factor for eVscopes\\
$M_{Unistellar}$ & ~~~~~~~~~~~~~~~~~~~~~~~~$\mathcal{N}$($0, 0.1)$ & Relative flux offset for eVscopes\\
$\sigma_{Unistellar}$ & ~~~~~~~~~~~~~~~~~~~~~~~~$\mathcal{LU}$($0.1, 10^5)$ & Jitter term for eVscope light curves (ppm)\\
$q_{1,Unistellar}$ & ~~~~~~~~~~~~~~~~~~~~~~~~0.4428 (fixed) & Quadratic limb darkening parametrization \citep{kipping2013efficient} \\
$q_{2,Unistellar}$ & ~~~~~~~~~~~~~~~~~~~~~~~~0.4135 (fixed) & Quadratic limb darkening parametrization \citep{kipping2013efficient} \\
\\\multicolumn{2}{l}{RV Parameters:}\smallskip\\
$\mu_{APF}$ & ~~~~~~~~~~~~~~~~~~~~~~~~$\mathcal{U}$($-1000, 1000)$ & Systemic velocity for APF (\ms) \\
$\sigma_{APF}$ & ~~~~~~~~~~~~~~~~~~~~~~~~$\mathcal{LU}$($10^{-3}, 1000)$ & Jitter term for APF (\ms) \\
$\mu_{CHIRON}$ & ~~~~~~~~~~~~~~~~~~~~~~~~$\mathcal{U}$($-10^5, 10^5)$ & Systemic velocity for CHIRON (\ms) \\
$\sigma_{CHIRON}$ & ~~~~~~~~~~~~~~~~~~~~~~~~$\mathcal{LU}$($10^{-3}, 10^5)$ & Jitter term for CHIRON (\ms) \\
$\mu_{TRES}$ & ~~~~~~~~~~~~~~~~~~~~~~~~$\mathcal{U}$($-1000, 1000)$ & Systemic velocity for TRES (\ms) \\
$\sigma_{TRES}$ & ~~~~~~~~~~~~~~~~~~~~~~~~$\mathcal{LU}$($10^{-3}, 1000)$ & Jitter term for TRES (\ms) \\
\label{tab:julietpriors} 
\enddata
\end{deluxetable*}

\startlongtable
\begin{deluxetable*}{llccc}
\tabletypesize{\small}
\tablewidth{\linewidth}
\tablecaption{Unistellar eVscope photometry posterior parameters (median values and 68\% confidence interval) from the \texttt{juliet} joint fit.}
\tablehead{\colhead{Parameter} & \colhead{Units} & \colhead{Values}}
\startdata
\\\multicolumn{2}{l}{eVscope-zzd}&\smallskip\\
$M$ & Relative flux offset & $0.0006$ \\
$\sigma$ &  Jitter term for light curve (ppm) & $15^{+418}_{-14}$ \\
\smallskip\\\multicolumn{2}{l}{eVscope-wt9}&\smallskip\\
$M$ & Relative flux offset & $-0.0085$ \\
$\sigma$ &  Jitter term for light curve (ppm) & $21^{+383}_{-21}$ \\
\smallskip\\\multicolumn{2}{l}{eVscope-wpg}&\smallskip\\
$M$ & Relative flux offset & $-0.0090$ \\
$\sigma$ &  Jitter term for light curve (ppm) & $7^{+209}_{-7}$ \\
\smallskip\\\multicolumn{2}{l}{eVscope-vxq}&\smallskip\\
$M$ & Relative flux offset & $0.0003$ \\
$\sigma$ &  Jitter term for light curve (ppm) & $17^{+440}_{-17}$ \\
\smallskip\\\multicolumn{2}{l}{eVscope-v8v}&\smallskip\\
$M$ & Relative flux offset & $-0.0040$ \\
$\sigma$ &  Jitter term for light curve (ppm) & $33^{+1075}_{-32}$ \\
\smallskip\\\multicolumn{2}{l}{eVscope-p6r}&\smallskip\\
$M$ & Relative flux offset & $-0.0082$ \\
$\sigma$ &  Jitter term for light curve (ppm) & $15^{+326}_{-14}$ \\
\smallskip\\\multicolumn{2}{l}{eVscope-kk2}&\smallskip\\
$M$ & Relative flux offset & $-0.0007$ \\
$\sigma$ &  Jitter term for light curve (ppm) & $14^{+277}_{-14}$ \\
\smallskip\\\multicolumn{2}{l}{eVscope-cxz}&\smallskip\\
$M$ & Relative flux offset & $-0.0001$ \\
$\sigma$ &  Jitter term for light curve (ppm) & $20^{+432}_{-20}$ \\
\smallskip\\\multicolumn{2}{l}{eVscope-cus}&\smallskip\\
$M$ & Relative flux offset & $-0.0077$ \\
$\sigma$ &  Jitter term for light curve (ppm) & $312^{+2568}_{-310}$ \\
\smallskip\\\multicolumn{2}{l}{eVscope-26e}&\smallskip\\
$M$ & Relative flux offset & $-0.0020$ \\
$\sigma$ &  Jitter term for light curve (ppm) & $14^{+509}_{-13}$ \\
\smallskip\\\multicolumn{2}{l}{eVscope-24s}&\smallskip\\
$M$ & Relative flux offset & $-0.0091$ \\
$\sigma$ &  Jitter term for light curve (ppm) & $22^{+499}_{-21}$ \\
\smallskip\\\multicolumn{2}{l}{eVscope-2rz}&\smallskip\\
$M$ & Relative flux offset & $0.0001$ \\
$\sigma$ &  Jitter term for light curve (ppm) & $25^{+479}_{-24}$ \\
\smallskip\\\multicolumn{2}{l}{eVscope-2n9}&\smallskip\\
$M$ & Relative flux offset & $-0.0089$ \\
$\sigma$ &  Jitter term for light curve (ppm) & $16^{+322}_{-16}$ \\
\label{tab:unistellarposteriors} 
\enddata
\end{deluxetable*}

\end{document}